%% file: paper.tex
\documentclass[a4paper,11pt]{article}

\usepackage{jcappub}
\usepackage{amsmath, amssymb}
\usepackage[sort&compress,numbers]{natbib}
\usepackage{txfonts}
\usepackage{siunitx}
\usepackage{graphicx}

\usepackage{color}
\usepackage{hyperref}
\usepackage{epstopdf}

\allowdisplaybreaks

\DeclareOldFontCommand{\bf}{\normalfont\bfseries}{\textbf}

\newcommand{\jchanges}[1]{#1}
\newcommand{\jchangesnew}[1]{#1}

\title{The Shape Dependence of Vainshtein Screening in the Cosmic Matter Bispectrum}

\author[a]{Clare Burrage,}
\author[a,1]{Johannes Dombrowski,\note{Corresponding author.}}
\author[a]{Daniela Saadeh}

\affiliation[a]{School of Physics and Astronomy, University of Nottingham,\\Nottingham, NG7 2RD, UK}

\emailAdd{Clare.Burrage@nottingham.ac.uk}
\emailAdd{Johannes.Dombrowski@nottingham.ac.uk}
\emailAdd{Daniela.Saadeh@nottingham.ac.uk}

\abstract{One of the most pressing questions in modified gravity is how deviations from general relativity can manifest in upcoming galaxy surveys. This is especially relevant for theories exhibiting Vainshtein screening, where such deviations are efficiently suppressed within a (typically large) Vainshtein radius. However, Vainshtein screening is known to be shape dependent: it is most effective around spherical sources, weaker around cylindrical objects and completely absent for planar sources. The Cosmic Web therefore offers a testing ground, as it displays many shapes in the form of clusters, filaments and walls.
	
In this work, we explicitly derive the signature of the shape dependence of Vainshtein screening on the matter bispectrum, by considering a cubic Galileon model with a conformal coupling to matter and a cosmological constant. We perform a second order perturbative analysis, deriving analytic, integral expressions for the bispectrum, which we integrate using \texttt{hi\_class}. We find that the shape dependence of Vainshtein screening enters the bispectrum with a unique scale-factor dependence of $\propto a^{3/2}$. The magnitude of the effect today is up to $\SI{2}{\percent}$ for a model whose linear growth rate deviates up to $\SI{5}{\percent}$ from $\Lambda$CDM.}


\arxivnumber{1905.06260}

\begin{document}
\maketitle
\flushbottom

\section{Introduction}

The ongoing search for an explanation of the value of the cosmological constant has motivated a remarkable amount of research in modified theories of gravity \cite{Bull,Joyce:2014kja}, in particular around the introduction of additional  light scalar degrees of freedom. Although no conclusive solution to the cosmological constant problem has been found, alternative theories of gravity nowadays have become a field of research in their own right. The question at hand is a much broader one: what is the fundamental nature of gravity? How could deviations from general relativity (GR) appear and be observed in current and future surveys?

An extension of GR via a light scalar field will introduce a long-range fifth force whose strength could be similar in magnitude to the gravitational force. Since constraints on fifth forces are tight on Solar System scales \cite{Will_review}, any significant modification of gravity must possess a screening mechanism suppressing the fifth force in the local environment. This can be obtained if non-linearities in the theory become important in specific regimes (for instance high density regions) whilst leaving the force unscreened on large scales.

One of the main contenders for such a screening mechanism is Vainshtein screening \cite{Vainshtein, BabichevVainshteinscreening} which suppresses the fifth force in regions of high second order derivatives of the scalar field. It was realized in Refs.~\cite{Brax:2011sv,Clare_shape_dependence} that the effectiveness of Vainshtein screening is strongly dependent on the shape of the source mass. The screening is most effective around spherical sources, less effective around cylindrical bodies and non-existent in systems with planar symmetry. 

The discovery of this shape-dependence motivates us to look for this effect in the cosmic web, where lots of different shapes are present, from clusters to filaments and walls. The dynamics of the cosmic web are investigated here using cosmological perturbation theory. Since Vainshtein screening is intrinsically non-linear, we have to go to at least second order in perturbation theory to observe any shape dependent effects: we will therefore use the matter bispectrum as our observable. The matter bispectrum is especially suitable to test for shape dependence as it is additionally sensitive to three wavenumbers, which form a closed triangle upon imposing background homogeneity. Different triangle shapes correspond to different shapes in real space, which can be more or less symmetric \jchanges{\cite{Lewis:2011au}}. 

The simplest model exhibiting Vainshtein screening is the cubic Galileon \cite{Nicolis, Deffayet2009_1, Deffayet2009_2}. Galileon fields are scalar fields that respect the Galilean shift symmetry $\partial_\mu\pi\rightarrow\partial_\mu \pi+a_\mu$ in flat space-time. Despite being characterised by higher derivative operators in the Lagrangian, the equation of motion of Galileon fields always remains second order, and therefore does not suffer from ghost instabilities. In four dimensions this Galileon model can be extended to include operators which have quartic and quintic dependence on the Galileon scalar field \cite{Nicolis}. It is also possible to extend the theory further if we just require second order equations of motion, but do not insist on the symmetry, to the so-called Horndeski scalar-tensor theory \cite{Horndeski:1974wa,Deffayet:2011gz}. In this work we restrict ourselves to the cubic Galileon as the simplest theory to exhibit the shape dependent behaviour of interest.

The Galileon was originally introduced as a possible explanation for the current  acceleration of the expansion of the universe without the need for a cosmological constant \cite{Nicolis}.  This was achieved through a mechanism known as self-acceleration \cite{Chow:2009fm,Silva:2009km,Gannouji:2010au}. Constraints on such self accelerating cosmologies come from observations of baryon acoustic oscillations, the cosmic microwave background, type 1a supernovae, the growth rate of structure, galaxy clustering and the integrated Sachs-Wolfe effect \cite{DeFelice:2010as,Ali:2010gr,Nesseris:2010pc,Hirano:2010yf,Hirano:2011wj,Appleby:2011aa,DeFelice:2011aa,Barreira:2012kk,Barreira:2013eea, Hellwing:2017pmj,Barreira:2013jma,Barreira:2014jha,Dinda:2018eyt}. Constraints also come from bounds on the time variation of the gravitational constant in the solar system \cite{Babichev:2011iz}, and variations of particle masses and fundamental constants \cite{Brax:2014vla,Brax:2015cla}.  For a general review we refer the reader to Refs.~
\cite{Khoury:2013tda,Joyce:2014kja,Koyama:2015vza,Ishak:2018his}. 

In the presence of the higher order Galileon operators, cosmological Galileon backgrounds also lead to differences between the speed of gravitational and electromagnetic waves \cite{Jimenez:2015bwa,Brax:2015dma}. These theories are now very tightly constrained by the recent observation of a neutron star merger simultaneously through gravitational and electromagnetic waves \cite{ZumaGWs, BakerGWs, SaksteinGWs, VernizziGWs,Crisostomi:2017pjs}. Thus, we restrict ourselves to cubic Galileons in this work.

It was shown in Ref.~\cite{RenkISW} that the cubic Galileon is ruled out as the only energy component driving the late-time accelerated expansion of the Universe, as it predicts a negative integrated Sachs-Wolfe effect in conflict with current obervations.  Therefore, we include a cosmological constant in our model and keep the Galileon energy density as a subdominant energy component throughout the entire evolution of the Universe. A cosmological model with such a sub-dominant cubic Galileon component was also considered in Ref.~\cite{Burrage:2015lla}, where it was shown that independent initial perturbations in the Galileon field can break the usual correlation between density and velocity  power spectra and lead to a form of stochastic bias.

The matter bispectrum was first studied for cubic Galileon models in Ref.~\cite{Matarrese} and, subsequently, generalised to Horndeski theories \cite{Japanese, Bellini2015, BAO_shift, Yamauchi:2017ibz}.
These analyses did not include an explicit coupling to matter, only an indirect coupling due to the Galileon and matter fields coupling to the metric. Our analysis includes an explicit conformal coupling of the Galileon field to matter, which results in a stronger fifth force and stronger effects from screening.

This paper is structured as follows. We begin by introducing our Galileon model and shortly reviewing the effects of shape dependence in Section \ref{Shape_dependence}. Then, we derive analytic expressions for the matter bispectrum using cosmological perturbation theory up to second order in Section \ref{Perturbation_theory}. In Section \ref{Numerics}, we evaluate the obtained integral expressions for the matter bispectrum numerically using the \jchangesnew{latest version of the} \texttt{hi\_class} code \jchangesnew{\cite{Bellini:2019syt, hi_class, Blas:2011rf}}, \jchangesnew{which is} capable of integrating the exact Galileon equation of motion. We then conclude in Section \ref{Conclusion}. Some of the more involved formulas and definitions are postponed to the appendices in order to guarantee a better flow of reading.

\section{Shape dependence of Vainshtein sceening}
\label{Shape_dependence}

In this section we introduce our Galileon model and review the effect of shape-dependent Vainshtein screening. As a simple proxy for an extension of general relativity with Vainshtein screening we consider a cubic Galileon model similar to the one used in Ref.~\cite{Matarrese}, but with a conformal coupling to the matter fields in order to have a stronger and more explicit fifth force. Furthermore, we include a cosmological constant $\Lambda$ since the cubic galileon cannot be the sole source of the late-time accelerated expansion of the Universe \cite{RenkISW}: 
\begin{equation}
S=\int\mathrm{d}^4x\left[ M_p^2\sqrt{-g}\left( \frac{R}{2}-\Lambda+\mathcal{L}_\pi \right) +\mathcal{L}_m\left[ \left(1+\pi\right)g_{\mu\nu}, \psi_i \right] \right], \label{action}
\end{equation}
with the Lagrangian for the Galileon fields being:
\begin{equation}
\mathcal{L}_\pi = -\frac{C_2}{2}(\nabla\pi)^2-\frac{C_3}{2}(\Box \pi)(\nabla\pi)^2,
\end{equation}
where $\Box\coloneqq\nabla_\mu\nabla^\mu$ with the covariant derivative $\nabla$. The exact form of the Lagrangian $\mathcal{L}_m$ of the matter fields $\psi_i$ is not important for the rest of the discussion; it is only relevant that the matter fields follow geodesics of the Jordan-frame metric $(1+\pi)g_{\mu\nu}$, which is called conformal coupling. In our conventions the field $\pi$ and the parameter $C_2$ are dimensionless and $C_3$ has inverse mass dimensions $2$. In the non-relativistic, weak-field limit the $(0,0)$-component of the Jordan-frame metric becomes $g_{00}=-(1+2\phi+\pi)$, with $\phi$ being the Newtonian gravitational potential. Particles following geodesics of the Jordan-frame metric are therefore subject to a fifth force mediated by the Galileon field:
\begin{equation}
\frac{\mathrm{d}^2\vec{x}}{\mathrm{d}t^2}= -\vec{\nabla} \phi +\vec{F}_5, \qquad \text{with} \qquad \vec{F}_5\coloneqq -\frac{\vec{\nabla} \pi}{2\jchanges{(1+\pi)}},
\end{equation}
where $\vec{\nabla}$ is the flat-space gradient in the 3 spatial directions.

Since we will use the \texttt{hi\_class} code for numerical analyses later on, we will use the \texttt{hi\_class} normalization for densities, such that the Friedmann equation takes the form $H^2= \sum_i \rho_i$ (the same normalization will be used for pressures). For the energy-momentum tensor of the matter fields we make the standard cosmological assumption that matter is non-relativistic can thus be treated as a pressureless, perfect fluid with velocity field $u_\mu$. With the \texttt{hi\_class} normalization of densities the energy-momentum tensor takes on the form $T^m_{\mu\nu}=3M^2_{p}\rho_m u_\mu u_\nu$. 

Variation of the action Eq.~\eqref{action} with respect to the Galileon field gives the field equation:
\begin{equation}
C_2\Box\pi+C_3\left( \left(\Box\pi\right)^2 - R_{\mu\nu}\nabla^\mu\pi\nabla^\nu\pi - \left(\nabla_\mu\nabla_\nu\pi\right)\left(\nabla^\mu\nabla^\nu\pi\right) \right)=\frac{3\rho_m}{2(1+\pi)}. \label{Gal_eom}
\end{equation}
Variation of the action with respect to the metric leads to the Einstein equations:
\begin{equation}
G_{\mu\nu}=M_p^{-2}\left( T_{\mu\nu}^{m} + T_{\mu\nu}^{\pi} \right) -\Lambda g_{\mu\nu}, \label{einstein}
\end{equation}
with the energy-momentum tensor of the Galileon field being:
\begin{eqnarray}
\frac{T_{\mu\nu}^\pi}{M^2_p}&=&C_2(\partial_\mu\pi) (\partial_\nu \pi) -\frac{C_2}{2}g_{\mu\nu}(\nabla\pi)^2 \nonumber \\
&&+C_3\left( (\partial_\mu\pi)(\partial_\nu\pi)\Box\pi -\nabla_{\left\{\mu\right.}\pi\nabla_{\left.\nu\right\}\alpha}\pi \nabla^{\alpha}\pi + g_{\mu\nu}\nabla^{\alpha}\pi\nabla_{\alpha\beta}\pi\nabla^{\beta}\pi \right).
\end{eqnarray}
Furthermore, the Bianchi identities give rise to the conservation equations:
\begin{equation}
\nabla^\mu \left( T^{m}_{\mu\nu} + T^{\pi}_{\mu\nu} \right)=0. \label{conservation}
\end{equation}

In order to demonstrate the shape dependence of the Vainshtein screening mechanism we examine Eq.~\eqref{Gal_eom} in the static and curvature free case and assume that $\pi\ll 1$:
\begin{equation}
C_2\Delta \pi +C_3\left( (\Delta\pi)^2-(\partial_i\partial_j\pi)(\partial^i\partial^j\pi) \right)=\frac{3\rho_m}{2}, \label{shape_term}
\end{equation}
where $\Delta$ is the static spatial Laplacian. This equation can be solved for various different shapes of the source $\rho_m$ \cite{Clare_shape_dependence}:
\begin{itemize}
	\item \textbf{Planar Symmetry:} \jchanges{For simplicity, we assume here that the source has the constant density $\rho_0$ if $z\leq z_0$ and zero otherwise. Any other configuration with planar symmetry qualitatively has the same result}. We consider the fifth force outside of the source and its relative strength with respect to the gravitational force $\vec{F}_G$:
	\begin{equation}
	\partial_z \pi =\frac{3\rho_0z_0}{2C_2}, \qquad \Rightarrow \qquad \frac{|\vec{F}_5|}{|\vec{F}_G|}=\frac{M_p^2}{2C_2}. \label{planar}
	\end{equation}
	For \jchanges{any configuration with planar symmetry} the non-linear ($C_3$) term in Eq.\ \eqref{shape_term} vanishes completely, there is no screening and the relative strength of the fifth force with respect to the gravitational force  is constant in space.
	\item \textbf{Cylindrical Symmetry:} We assume the source to have constant density $\rho_0$ inside the radius $r_0$ (radius being defined as $r^2=x^2+y^2$ here) and zero otherwise:
	\begin{equation}
	\partial_r\pi=\frac{C_2r}{2C_3}\left( \sqrt{1+\frac{r_v^2}{r^2}}-1 \right) \qquad \text{with} \qquad r_v\coloneqq \sqrt{\frac{3r_0^2\rho_0C_3}{C_2^2}}.
	\end{equation}
	$r_v$ is called the Vainshtein radius and determines the scale below which the non-linear terms become important. Within the Vainshtein radius ($r_0<r\ll r_v$):
	\begin{equation}
	\frac{|\vec{F}_5|}{|\vec{F}_G|}= \frac{M_p^2}{C_2}\frac{r}{r_v}.
	\end{equation}
	We see that the fifth force becomes weaker than the graviational force inside the Vainshtein radius and is thus screened.
	\item \textbf{Spherical Symmetry:} The source is defined in the same way as for the cylindrical case except now radii are defined by $r^2=x^2+y^2+z^2$. We obtain:
	\begin{equation}
	\partial_r\pi=\frac{C_2r}{4C_3}\left( \sqrt{1+\frac{r_v^3}{r^3}} -1 \right) \qquad \text{with} \qquad r_v\coloneqq \left( \frac{4r_0^3\rho_0C_3}{C_2^2} \right)^{1/3}. \label{spherical_symm}
	\end{equation}
	The relative strength of the fifth force inside the Vainshtein radius is now given by:
	\begin{equation}
	\frac{|\vec{F}_5|}{|\vec{F}_G|}= \frac{M_p^2}{C_2}\frac{r^{3/2}}{r_v^{3/2}}. \label{spherical}
	\end{equation}
	We observe that the fifth force is screened more effectively inside the Vainshtein radius than for the cylindrical case.
\end{itemize}
To summarize, the more evenly the Galileon field depends on the three directions of space, the larger is the non-linear term in Eq.\ \eqref{shape_term} and the more effective is Vainshtein screening. All of the above mentioned source symmetries appear on cosmological scales as walls, filaments and halos. Therefore, we will search for this shape-dependent effect in the distribution of cosmic structure.

More specifically, the shape-dependent term in Eq.\ \eqref{shape_term} is non-linear, so that its effects can only manifest at second or higher order in cosmological perturbation theory. As a proxy for the shape dependence, we will therefore consider the matter bispectrum, which is sensitive to both non-linearities and shapes, due to its dependence on three wavevectors forming a closed triangle. Based on Ref.~\cite{Clare_shape_dependence}, we expect both the non-linearities and screening to be most effective for the equilateral configuration, which corresponds to the highest degree of symmetry, and least effective for the flattened configuration, where all three sides of the triangle are parallel thus containing only one-dimensional information in Fourier space.

\section{Cosmological perturbation theory in the Einstein frame}
\label{Perturbation_theory}

In order to compute the bispectrum for our Galileon model, we have to perform a perturbative analysis up to second order in the matter density contrast. For this we start by analysing the background evolution and then proceed with a first and second order calculation. A similar analysis without an explicit coupling to matter was performed in Ref.\ \cite{Matarrese}.

\subsection{Background evolution}

At background level, the metric is given by the FLRW-metric. We choose to work with conformal time $\tau$:
\begin{equation}
\mathrm{d}s^2=a(\tau)^2\left( -\mathrm{d}\tau^2 +\delta_{ij}\mathrm{d}x^i\mathrm{d}x^j \right).
\end{equation}
In the following, derivatives with respect to conformal time will be denoted by a prime. From this point on we will set $M_p=1$. Defining the conformal Hubble function $\mathcal{H}=a'/a$, the Einstein equations become:
\begin{eqnarray}
&\frac{\mathcal{H}^2}{a^2}=\rho_m+\rho_\pi +\frac{\Lambda_c}{3}, \label{Friedmann_Einstein_frame} \\
&\frac{1}{a^2}\left( \mathcal{H}^2+2\mathcal{H}' \right)=-p_\pi+\frac{\Lambda_c}{3},
\end{eqnarray}
where the background Galileon density and pressure are defined by:
\begin{eqnarray}
\rho_\pi&=& \frac{C_2}{6a^2}\pi'^2 -\frac{C_3}{a^4}\mathcal{H}\pi'^3, \label{Gal_dens_back} \\
p_\pi&=& \frac{C_2}{6a^2}\pi'^2 + \frac{C_3}{3a^4}\pi'^2\left( \pi''-\mathcal{H}\pi' \right).
\end{eqnarray}
The background Galileon equation of motion takes on the form:
\begin{equation}
-\frac{C_2}{a^2}\left( \pi'' + 2\mathcal{H}\pi' \right) + \frac{3C_3}{a^4}\pi'\left( 2\mathcal{H}\pi'' + \mathcal{H}'\pi' \right)= \frac{3\rho_m}{2(1+\pi)}. \label{Galileon_background}
\end{equation}
Finally, the 0-component of the conservation equations becomes:
\begin{equation}
-\rho_m'-3\mathcal{H}\rho_m +\frac{\rho_m \pi'}{2(1+\pi)}=0. \label{cont_background}
\end{equation}

\subsection{Cosmological Vainshtein screening}
\label{Vainshtein_cosmo}

Before we solve the Galileon field equation \eqref{Galileon_background} numerically with \texttt{hi\_class}, we can obtain useful analytical understanding by applying the simplifying assumption $\pi \ll 1$ which is typically true if the Galileon density \eqref{Gal_dens_back} is subdominant on the background. This assumption enables us to approximate the continuity equation \eqref{cont_background} by $\rho_m'=-3\mathcal{H}\rho_m$ if we additionally assume the quasi-static approximation $\pi'\sim \mathcal{H}\pi$. Both assumptions have been checked numerically for all the models we study in Section \ref{Numerics}, and the quasi static approximation has been previously shown to hold for theories with Vainshtein screening in Ref.~\cite{Winther:2015pta}. Under these assumptions the continuity equation has the solution $\rho_m=\rho_{m, 0} a^{-3}$. It is now straightforward to solve the Galileon equation of motion:
\begin{equation}
\pi'_{1/2}=\frac{C_2a^2}{6C_3\mathcal{H}}\left( 1\pm \sqrt{1+\lambda_v(a)} \right), \quad \text{with} \quad \lambda_v(a)\coloneqq 18H\rho_m t\frac{C_3}{C_2^2}, \label{pi_prime}
\end{equation}
where $t$ is the physical time. Variants of this solution, which assumed $C_2$ to be negligibly small, have already been derived in Refs.~\cite{Chow:2009fm,Burrage:2015lla}. By solving the equation of motion numerically using \texttt{hi\_class} we found for all models studied in Section \ref{Numerics} that only the negative branch of the solution is stable \jchanges{on a cosmological background}. 

The solution \eqref{pi_prime} is called the cosmological Vainshtein solution, because it exhibits a time-like Vainshtein screening effect. The function $\lambda_v$ measures the magnitude of the non-linear terms in the equation of motion against the linear terms and is the time-like equivalent of $(r_v/r)^3$ in the spatial solution around a spherical source \eqref{spherical_symm}. We will show in the next Section that the linearised fifth force will be screened at early times where $\lambda_v\gg 1$ compared to a situation where $\lambda_v\ll 1$.

\subsection{Linear Perturbation Theory}

We now look at scalar linear perturbations in the cosmic fluid. We choose Newtonian gauge such that the perturbed metric can be written in the following form:
\begin{equation}
\mathrm{d}s^2=a^2\left[ -(1+2\psi)\mathrm{d}\tau^2 +(1-2\phi)\delta_{ij}\mathrm{d}x^i\mathrm{d}x^j \right]. \label{pert_metric}
\end{equation}
We parametrize the matter and velocity field perturbations by:
\begin{equation}
\rho_m=\bar{\rho}(1+\delta), \quad u^\mu=\frac{1}{a}\left[ \delta^\mu_0+v^\mu \right], \quad v^\mu= (v^0, \nabla v)^\top. \label{perturbative_decomposition}
\end{equation}
The perturbed Galileon field is written as $\pi+\delta \pi$, where $\pi$ is the background field and $\delta\pi$ is the perturbative variable. The 0-component of the velocity, $v^0$, is determined by the constraint $u^\mu u_\mu=-1$. The perturbation variables $\phi$, $\delta$, $\delta\pi$, \dots are expanded into first order, second order, etc.\ perturbations, for example:
\begin{equation}
\delta\pi=\pi^{(1)}+\frac{\pi^{(2)}}{2}+\dots
\end{equation}
The Galileon field, Einstein and Conservation equations \eqref{Gal_eom}, \eqref{einstein}, \eqref{conservation} at linear order are listed in \ref{appendix}. Combining these equations in Fourier-space using the quasi-static approximation, which we already used in the previous Section \ref{Vainshtein_cosmo}, and assuming subhorizon modes only ($\mathcal{H}^2\ll k^2$) we get a second order differential equation for the density contrast:
\begin{eqnarray}
\delta^{(1)\prime\prime}+\left( \mathcal{H}+\frac{\pi'}{2(1+\pi)} \right)\delta^{(1)\prime}= \alpha_\delta(\tau)\delta^{(1)}, \label{lin_pert}
\end{eqnarray}
where
\begin{eqnarray}
\alpha_\delta&\coloneqq& \frac{3a^2\bar{\rho}}{2}\left[ 1+\frac{\alpha_\pi^2}{2A(\tau)} \right], \quad \alpha_\pi\coloneqq \frac{1}{1+\pi}+\frac{C_3\pi'^2}{a^2}, \\
A(\tau)&\coloneqq&C_2-\frac{2C_3}{a^2}(\pi''+\mathcal{H}\pi')-\frac{C_3^2\pi'^4}{2a^4}. \label{A_Def}
\end{eqnarray}
We see that the gravitational force of strength $M_p^{-2}=1$ is accompanied by a fifth force of strength $\alpha_\pi^2/(2A)$. The first term in $\alpha_\pi$ reflects the conformal coupling between matter and the Galileon field, and the second term is a consequence of matter coupling to the Galileon field indirectly with gravity as mediator. Inserting the cosmological Vainshtein solution \eqref{pi_prime} into the expression for $A(\tau)$ enables us to observe the cosmological screening of the linearised fifth force. Assuming matter domination for simplicity (radiation domination just gives different numerical factors), $A(\tau)=C_2(1+\lambda_v(a)/3)\approx C_2$ if $\lambda_v(a)\ll 1$ and $A=C_2(1+2\sqrt{\lambda_v}/3)\approx 2C_2\sqrt{\lambda_v}/3$ if $\lambda_v\gg 1$. We conclude that for $\lambda_v\ll 1$ the fifth force is unscreened with strength $\sim C_2^{-1}$ relative to the gravitational force which is analogous to the result \eqref{planar}, and for $\lambda_v\gg 1$ the fifth force is screened with strength $(C_2\sqrt{\lambda_v})^{-1}$ relative to the gravitational force, with $\lambda_v$ being the equivalent of $(r_v/r)^3$.

\subsection{Breakdown of Perturbation Theory}
\label{sec:condition}

In general relativity, cosmological perturbation theory is restricted to scales that obey $\delta\ll 1$. Below the non-linearity scale, \jchanges{the hierarchy between background, linear and non-linear terms in the Continuity and Euler equations breaks down. Similarly, for the background Galileon equation of motion \eqref{Galileon_background} we had to assume $\left|\Box\delta\pi(\tau, \vec{x})\right|\ll\left|\Box \pi(\tau)\right|$, i.e.~we neglected second derivatives of the spatially dependent part $\delta\pi$ of the Galileon field compared to second derivatives of the background field $\pi$ . This means that there is a second non-linearity scale in our system, the scale where $\left|\Box\delta\pi(\tau, \vec{x})\right|\sim\left|\Box \pi(\tau)\right|$ and a perturbative analysis breaks down.} 

\jchanges{A similar assumption had to be made when we derived the linear equation of motion \eqref{Galileon_pre}. Additionally to assuming $\delta\pi\approx\pi^{(1)}$, we had to assume that terms non-linear in $\delta\pi$, like $C_3(\Box\delta\pi)^2$, are small compared to the linear term $C_3\Box\pi(\tau)\Box\delta\pi$. This makes the connection between this new linearity scale and the Vainhstein radius clear as both indicate the scale at which the non-linearities of the spatially dependent Galileon field become relevant in the equation of motion.}

\jchanges{If this new non-linearity scale is larger than the standard non-linearity scale defined by $\delta\sim 1$, our analysis is even more restricted than a conventional perturbative analysis without the Galileon field. Therefore, it is important to check how the assumption
\begin{equation}
\left|\frac{\Box\delta\pi(\tau, \vec{x})}{\Box \pi(\tau)}\right|\ll 1 \label{condition_1} 
\end{equation}
compares to the assumption $\delta\ll 1$. The spatially dependent field $\delta\pi$ on the left-hand side of the condition \eqref{condition_1} is in principle given by the full series $\delta\pi=\sum_{n=1}^\infty\pi^{(n)}/n!$ making the condition a non-perturbative statement about the perturbativity of the system. However, this makes it unfeasible to check the condition in practice. We therefore assume in the following that the order of magnitude of the non-linearity scale can be estimated if we simplify $\delta\pi\approx\pi^{(1)}$. This simplification is true on all scales up to the non-linearity scale, so the estimate for the non-linearity scale obtained in this way should be of the correct order of magnitude. Furthermore, we assume that the quasi-static approximation is valid, i.e.~we can neglect the time derivatives in $\Box\delta\pi$ compared to the spatial derivatives. This assumption is valid as long as the non-linearity scale is a sub-horizon scale. With these assumptions the ratio on the left-hand side of the condition \eqref{condition_1} is given by:
\begin{equation}
\frac{\Delta\pi^{(1)}}{(\pi^{\prime\prime}+2\mathcal{H}\pi')}= \frac{3\bar{\rho}\alpha_\pi a^2}{2A(\tau)(\pi^{\prime\prime}+2\mathcal{H}\pi')}\delta^{(1)}.
\end{equation}
We used that, on a linear level, the Einstein and Galileon field equations \eqref{Poisson} and \eqref{Galileon_pre} can be combined to:
\begin{eqnarray}
\Delta \pi^{(1)} = \frac{3\bar{\rho}\alpha_\pi}{2A(\tau)}\delta^{(1)}. \label{pi_first_order}
\end{eqnarray}
We again made use of the quasi-static assumption. This shows that as long as:
\begin{equation}
\lambda(\tau)\coloneqq\left|\frac{2A(\tau)(\pi^{\prime\prime}+2\mathcal{H}\pi')}{3\bar{\rho}\alpha_\pi a^2}\right|\gtrsim 1, \label{condition}
\end{equation}
the new non-linearity scale leads to no additional restrictions of the perturbative analysis.}

We will check this condition for every model we study in Section \ref{Numerics} and plot the quantity $\lambda$ as a function of time in Figure \ref{fig:perturbation}. \jchanges{In a simplified, analytically tractable setting, where we assume matter domination together with $C_2=0$, we find $\lambda=2$, i.e.~the two non-linearity scales are of the same order of magnitude. What appears to be a coincidence at first sight is actually a consequence of the particular Galileon model we chose. The conformal coupling of the Galileon field with matter causes the background equation of motion to require $C_3(\Box\pi)^2\sim\bar{\rho}$ and the linear equation to enforce $C_3\Box\pi\Box\delta\pi\sim\bar{\rho}\delta$. Combining these two scaling relations, results in $\left|\frac{\Box\delta\pi(\tau, \vec{x})}{\Box \pi(\tau)}\right|\sim\delta$. Since it is not trivial that the two non-linearity scales will be equivalent for any Vainshtein screening model, we propose that the condition \eqref{condition_1} should be checked for any perturbative study of Vainshtein screening.}

\subsection{Second Order Perturbation Theory}
\label{sec:second_order}

In order to compute the matter bispectrum we have to proceed to second order in perturbation theory which enables us to capture the onset of non-linear dynamics and the Vainshtein screening mechanism. Computing the Galileon field, Einstein and Conservation equations Eqs.~\eqref{Gal_eom}, \eqref{einstein}, \eqref{conservation} at second order and combining them appropriately, results in the inhomogeneous equation:
\begin{eqnarray}
\delta^{(2)\prime\prime}+\left( \mathcal{H}+\frac{\pi'}{2(1+\pi)} \right)\delta^{(2)\prime}= \alpha_\delta(\tau)\delta^{(2)}+ S^{(\delta)}, \label{density_contrast2}
\end{eqnarray}
where the inhomogeneity $S^{(\delta)}$ captures the non-linear physics. In \ref{sec:source_term} we state the source function $S^{(\delta)}$ in terms of the source functions $S^{(1)}$, $S^{(4)}$, $S^{(5)}$, $S^{(6)}$ and $S^{(7)}$ which were defined in the Appendix of Ref.~\cite{Matarrese} and are sufficiently long that we don't reproduce them here. However, we would like to point out the appearance of one crucial term, which was also found in Ref.~\cite{Japanese}: 
\begin{equation}
S^{(\delta)}\supset -\frac{C_3\alpha_\pi}{a^2A(\tau)}\left[ \left( \Delta\pi^{(1)} \right)^2 - \left( \partial_i\partial_j\pi^{(1)} \right)\left( \partial^{i}\partial^j\pi^{(1)} \right) \right]. \label{S_delta_shape}
\end{equation}
This terms has the same structure as the non-linear part of Eq.\ \eqref{shape_term} and thus encodes the shape dependence of the Vainshtein screening mechanism. 

Making use of the quasi-static approximation and a Fourier transform, $S^{(\delta)}$ can be brought into the form:
\begin{equation}
S^{(\delta)}=\int\mathrm{d}^3k_1\mathrm{d}^3k_2\,\delta(\vec{k}-\vec{k}_1-\vec{k}_2)\,\mathcal{K}(a, \vec{k}_1, \vec{k}_2)\,\delta^{(1)}(a, \vec{k}_1)\,\delta^{(1)}(a, \vec{k}_2). \label{kernel_def}
\end{equation}
where the kernel $\mathcal{K}$ is given by:
\begin{eqnarray}
\mathcal{K}(a, \vec{k}_1, \vec{k}_2)&=& 2\left[ \left(1+\alpha(\vec{k}_1, \vec{k}_2)\right) \left( \mathcal{H}^2f^2(\tau) +\alpha_\delta(\tau) \right) \right. \nonumber \\
& &\left. -\frac{9C_3\bar{\rho}^2a^2\alpha_\pi^3}{8A^3(\tau)}  \gamma(\vec{k}_1, \vec{k}_2) +\beta(\vec{k}_1, \vec{k}_2)\mathcal{H}^2f^2(\tau) \right], \label{kernel}
\end{eqnarray}
with the linear growth rate $f=\mathrm{d}\ln D_+/\mathrm{d}\ln a$, where $D_+$ is the linear growth factor. We also made use of the first order result in Eq.~\eqref{pi_first_order} in order to express $\pi^{(1)}$ in terms of the density contrast $\delta^{(1)}$. The $k$-dependencies are captured by the form factors:
\begin{eqnarray}
\alpha(\vec{k}_1, \vec{k}_2)&\coloneqq & \frac{\vec{k}_1\cdot\vec{k}_2}{2k_1^2k_2^2}(k_1^2+k_2^2), \\
\beta(\vec{k}_1, \vec{k}_2)& \coloneqq & \frac{\vec{k}_1\cdot\vec{k}_2(\vec{k}_1+\vec{k}_2)^2}{2k_1^2k_2^2}, \\
\gamma(\vec{k}_1, \vec{k}_2)&\coloneqq & 1-\frac{(\vec{k}_1\cdot\vec{k}_2)^2}{k_1^2k_2^2}.
\end{eqnarray}
The form factors $\alpha$ and $\beta$ are standard form factors appearing in cosmological perturbation theory in GR \cite{Bernardeau}. The additional form factor $\gamma$ originates directly as a Fourier transform of \eqref{S_delta_shape} and reflects very intuitively the shape dependence of Vainshtein screening. If the modes $\vec{k}_1$ and $\vec{k}_2$ are parallel, the term vanishes and no screening can occur. $\vec{k}_1$ and $\vec{k}_2$ being parallel means that we only capture one-dimensional information which is equivalent to a situation in real space with planar symmetry where, similarly  no screening occurs, compare with Eq.\ \eqref{planar}. The appearance of $\gamma$ in Horndeski theories was already noted in Refs.~\cite{Japanese, Yamauchi:2017ibz}, but without making the connection to the shape dependence of Vainshtein screening.

Making the ansatz:
\begin{equation}
\delta^{(2)}(\tau, \vec{k})=\int\mathrm{d}^3k_1\mathrm{d}^3k_2\,\delta(\vec{k}-\vec{k}_1-\vec{k}_2)\,F_2(\tau, \vec{k}_1, \vec{k}_2)\,\delta^{(1)}(\tau, \vec{k}_1)\,\delta^{(1)}(\tau, \vec{k}_2),
\end{equation}
we can solve the inhomogeneous differential equation \eqref{density_contrast2} with Green's method:
\begin{equation}
F_2(\tau, \vec{k}_1, \vec{k}_2)=\int_{\tau_i}^{\tau}\mathrm{d}\tilde{\tau}\,G(\tau, \tilde{\tau})\,\mathcal{K}(\tilde{\tau}, \vec{k}_1, \vec{k}_2) \frac{D_+^2(\tilde{\tau})}{D_+^2(\tau)}. \label{form_factor}
\end{equation}
The Green's function $G$ is defined by:
\begin{equation}
G(\tau, \tilde{\tau})\coloneqq\frac{D_1(\tau)D_2(\tilde{\tau})- D_2(\tau)D_1(\tilde{\tau})}{W(\tilde{\tau})}\Theta(\tau-\tilde{\tau}), \label{Greens_function}
\end{equation}
with the Wronskian $W$:
\begin{equation}
W(\tau)\coloneqq D_1'(\tau)D_2(\tau)-D_2'(\tau)D_1(\tau). \label{Wronskian}
\end{equation}
The functions $D_1$ and $D_2$ are two independent solutions of the linear growth equation \eqref{lin_pert}. In our numerical analysis in Section \ref{sec:numerics_linear} we will associate $D_1$ and $D_2$ with the growing and decaying modes $D_+$ and $D_-$.

Assuming Gaussian initial conditions, the form factor $F_2$ is directly connected to the bispectrum:
\begin{equation}
B(\tau, \vec{k}_1, \vec{k}_2, \vec{k}_3)=F_2(\tau, \vec{k}_1, \vec{k}_2)P(\tau, k_1)P(\tau, k_2)+\text{cycl. Perm.}, \label{bispectrum}
\end{equation}
where $P(\tau, k)$ is the linearly evolved power spectrum. Therefore, the bispectrum depends on both the linear perturbations and the second order perturbations. It is thus convenient to introduce the reduced bispectrum:
\begin{equation}
Q(\tau, \vec{k}_1, \vec{k}_2, \vec{k}_3) \coloneqq \frac{B(\tau, \vec{k}_1, \vec{k}_2, \vec{k}_3)}{P(\tau, k_1)P(\tau, k_2)+\text{cycl. perm.}}, \label{red_bispec}
\end{equation}
where the linear growth of the power spectra cancels out. Furthermore, the reduced bispectrum has the advantage of being mostly scale independent \cite{Bernardeau}.

Using the kernel in Eq.~\eqref{kernel}, $F_2$ can be cast into the form:
\begin{equation}
F_2(\tau, \vec{k}_1, \vec{k}_2)=\mathcal{A}_{GR}(\tau)\left( 1+\alpha(\vec{k}_1, \vec{k}_2) \right) +\mathcal{B}_{GR}(\tau)\,\beta(\vec{k}_1, \vec{k}_2)+\mathcal{B}_\pi(\tau)\,\gamma(\vec{k}_1, \vec{k}_2). \label{F2_extended}
\end{equation} 
The time dependent functions $\mathcal{A}_{GR}$, $\mathcal{B}_{GR}$ and $\mathcal{B}_\pi$ are defined in \ref{appendix}. While the functions $\mathcal{A}_{GR}$ and $\mathcal{B}_{GR}$ always appear in general relativity, $\mathcal{B}_\pi$ is a purely Galileon contribution describing the shape-dependent non-linearities in the Galileon equation of motion Eq.~\eqref{shape_term}. We demonstrate in \ref{sec:F_2_simplification} that there is a relation between $\mathcal{A}_{GR}$ and $\mathcal{B}_{GR}$\footnote{We thank Emilio Bellini for pointing out this relation to us.}:
\begin{equation}
\mathcal{A}_{GR}(\tau)= 2-\mathcal{B}_{GR}(\tau).
\end{equation}
Defining $\mu$ as the cosine of the angle between $\vec{k}_1$ and $\vec{k}_2$, we thus conclude:
\begin{equation}
F_2(\tau, \vec{k}_1, \vec{k}_2)=2+\frac{\mu}{k_1k_2}\left( k_1^2+k_2^2 \right) - \mathcal{B}(\tau)\left( 1-\mu^2 \right), \label{F_2}
\end{equation}
where
\begin{equation}
\mathcal{B}(\tau)\coloneqq\mathcal{B}_{GR}(\tau)-\mathcal{B}_\pi(\tau)
\end{equation}
So we see that the shape-dependence of Vainshtein screening enters the matter bispectrum as a correction to the GR contribution $\mathcal{B}_{GR}$. The origin of $\mathcal{B}_{GR}$ can be traced back to non-linearities in the continuity and Euler equations where the flow of matter enters the total time derivative $\mathrm{d}/\mathrm{d}t=\partial/\partial t+\vec{v}\cdot\nabla$.

The contributions from Vainshtein screening to the form factor $F_2$, i.e. the term $\mathcal{B}_\pi(\tau)(1-\mu^2)$,
vanish in the flattened limit, where all the three wavevectors are approximately parallel and $\mu^2 \approx 1$. In this case, the real-space, plane-wave density perturbations associated with these three wavevectors will only depend on one direction of space which is equivalent to the planar symmetry discussed in Section \ref{Shape_dependence}.

The nature of the signal imprinted by Vainshtein screening on the bispectrum may seem counter-intuitive: it vanishes where the fifth force is unscreened and is maximum where the screening is also largest. This is because to stronger screening correspond larger non-linearities, which are detected by the bispectrum. A situation of no screening corresponds to no additional non-linearities.

It is possible to compute the function $\mathcal{B}_\pi$ analytically in a simplified setting. During the rest of this Section we assume matter domination and that $C_2=0$. During matter domination, $D_+\propto a$ and $D_-\propto a^{-3/2}$ with $a=\rho_{m, 0} \tau^2/4$, so the Green's function takes on the form:
\begin{equation}
G(\tau, \tilde{\tau}) = \frac{1}{5}\left( \frac{\tau^2}{\tilde{\tau}}- \frac{\tilde{\tau}^4}{\tau^3} \right).
\end{equation}
The negative (stable) branch of the solution to the Galileon field equation Eq.~\eqref{pi_prime} becomes:
\begin{equation}
\pi'= -\sqrt{\frac{\rho_{m, 0}t}{2C_3\mathcal{H}}}=-\frac{a}{\sqrt{3C_3}} \quad \Rightarrow \quad \pi= \frac{2\pi'}{3\mathcal{H}} \quad \text{and} \quad \pi^{\prime\prime} = \mathcal{H}\pi'.
\end{equation}
This enables us to compute:
\begin{equation}
\alpha_\pi = \frac{4}{3}, \qquad A(\tau)= -\frac{8}{9\pi}-\frac{1}{18}\approx -\frac{8}{9\pi},
\end{equation}
where we assumed $\pi\ll 1$. Using Eq.~\eqref{Gal_dens_back} we find $\rho_\pi/\rho_m=-\pi/2$ and we conclude:
\begin{equation}
-\frac{9C_3\bar{\rho}^2a^2\alpha_\pi^3}{8A^3(\tau)} = \frac{9}{16}\mathcal{H}^2\pi = -\frac{9}{8}\mathcal{H}^2\frac{\rho_\pi}{\rho_m}.
\end{equation}
Finally, we can integrate for $\mathcal{B_\pi}$:
\begin{eqnarray}
\mathcal{B_\pi}= -2\int_{\tau_i}^{\tau}\mathrm{d}\tilde{\tau}\,G(\tau, \tilde{\tau}
)\frac{9C_3\bar{\rho}^2a^2\alpha_\pi^3}{8A^3(\tau)}\frac{D_+^2(\tilde{\tau})}{D_+^2(\tau)} = -\frac{9}{50}\frac{\rho_\pi}{\rho_m}. \label{analytic}
\end{eqnarray}
The fraction $\rho_\pi/\rho_m$ is small due to the assumption of matter domination. During matter domination, $\mathcal{B}_{GR}$ has the standard value of $4/7$, see e.g.\ Ref.~\cite{Bernardeau}, with small corrections given by the impact of the Galileon field on the background evolution. These corrections are difficult to compute analytically but will be studied numerically in Section \ref{sec:numerics_bispectrum}. We see that the contributions $\mathcal{B}_\pi$ from the Galileon field are small compared to $\mathcal{B}_{GR}$, but in contrast to $\mathcal{B}_{GR}$ evolve in time, which helps break the degeneracy between the two terms. 

\section{Numerical analysis with \texttt{hi\_class}}
\label{Numerics}

In order to back up our analytic approximation in Eq.~\eqref{analytic}, and to generalise it beyond matter domination, we will evaluate the functions $\mathcal{B}_{GR}$ and $\mathcal{B}_\pi$ with \texttt{hi\_class}\footnote{\href{www.hiclass-code.net}{www.hiclass-code.net}} \jchangesnew{\cite{Bellini:2019syt, hi_class}}, a Boltzmann solver for Horndeski-type models \jchangesnew{based on CLASS \cite{Blas:2011rf}}. \jchangesnew{While previous public versions of} \texttt{hi\_class} \jchangesnew{\cite{hi_class}} require\jchangesnew{d} one to parameterise the time evolution of the Horndeski $\alpha$ functions as defined in Ref.~\cite{alpha_functions} in order to fully evolve the system\jchangesnew{, the latest version of the code \cite{Bellini:2019syt} is able to integrate the full equation of motion of any Horndeski theory, including the Galileon}.  

Since \texttt{hi\_class} works in the Jordan frame, we have to transform our action, Eq.\ \eqref{action}, with the conformal transformation 
\begin{equation}
g_{\mu\nu} \rightarrow \tilde{g}_{\mu\nu}= \Omega^2 g_{\mu\nu}, \quad\text{where} \quad \Omega^2\coloneqq 1+\pi. \label{transformation}
\end{equation}
This transformation brings our model into the Jordan frame and allows it to be formulated in terms of the standard Horndeski functions $G_i(\pi, X)$ \cite{Horndeski:1974wa,Deffayet:2011gz}. The form of the Horndeski functions for our model is shown in \ref{appendix}.

When working with both the Jordan and Einstein frame, one has to make sure to connect correctly between physical quantities in both frames. More specifically, we have to consider the density contrast in Einstein ($\delta$) and Jordan ($\tilde{\delta}$) frame which are related by:
\begin{eqnarray}
\tilde{\delta}^{(1)}=\delta^{(1)}-2\frac{\pi^{(1)}}{1+\pi}.
\end{eqnarray}
Using Eq.~\eqref{pi_first_order}, we can show that:
\begin{equation}
\tilde{\delta}^{(1)}-\delta^{(1)}= -2\frac{\pi^{(1)}}{1+\pi}\sim \frac{\mathcal{H}^2}{k^2}\delta^{(1)}.
\end{equation}
Therefore, on subhorizon scales and assuming the quasi-static approximation, the density contrast becomes approximately the same in both frames, see also Ref\jchanges{s}.\ \jchanges{\cite{Baojiu, Francfort:2019ynz}}. 
As we are only interested in the subhorizon scales where non-linear dynamics become important, the density contrast is effectively invariant under the conformal transformation and so is the matter bispectrum.

Other quantities like the Hubble function $\mathcal{H}$, the matter density $\rho_m$, the scale factor $a$, have to be transformed carefully under the conformal transformation. A summary of these transformations is given in \ref{appendix}.

\subsection{Background evolution}

We now analyse the background evolution of the Universe in our Galileon model numerically using \texttt{hi\_class}. 

There is a subtlety with regards to the Galileon density $\rho_\pi$ when working in the Jordan frame, because the structure of the Friedmann equation \eqref{Friedmann_Einstein_frame} changes significantly. In the Jordan frame we have:
\begin{eqnarray}
& &\frac{1}{a^2(1+\pi)}\left( \mathcal{H}^2 -\frac{\mathcal{H}\pi'}{1+\pi}+\frac{\pi^{\prime 2}}{4(1+\pi)^2} \right) \nonumber \\
& & \qquad\qquad=\frac{1}{1+\pi}\frac{C_2 \pi^{\prime 2}}{6a^2} +\frac{C_3\pi^{\prime 3}}{a^4}\left( \mathcal{H}-\frac{\pi'}{2(1+\pi)} \right) + \rho_m + \frac{\Lambda}{3(1+\pi)^2}. \label{Friedmann_Jordan}
\end{eqnarray}
This equation can be cast into the traditional form of a Friedmann equation, which is assumed by \texttt{hi\_class}, by defining an effective Galileon energy density:
\begin{equation}
\frac{\mathcal{H}^2}{a^2}=\rho_m+\rho_{\pi, \text{eff}} +\frac{\Lambda}{3},
\end{equation}
where
\begin{eqnarray}
\rho_{\pi, \text{eff}} &\coloneqq& \frac{1}{1+\pi}\frac{C_2 \pi^{\prime 2}}{6a^2} +\frac{C_3\pi^{\prime 3}}{a^4}\left( \mathcal{H}-\frac{\pi'}{2(1+\pi)} \right) \nonumber \\
& &+ \frac{1}{a^2(1+\pi)}\left( \mathcal{H}^2\pi + \frac{\mathcal{H}\pi'}{1+\pi}-\frac{\pi'^2}{4(1+\pi)^2}-\Lambda\frac{2\pi+\pi^2}{3(1+\pi)} \right). \label{effective_density}
\end{eqnarray}
This effective Galileon density measures all deviations from a $\Lambda$CDM cosmology. Similarly, an effective Galileon pressure $p_{\pi, \text{eff}}$ can be defined. 

The \texttt{hi\_class} code checks for instabilities of the background by calculating the sign of the kinetic term and the sound speed of the scalar field. For all our models it turns out that stability is guaranteed if the Galileon field is negative. As a consequence, $\rho_{\pi,\text{eff}}$ is negative as well. However, the physical Galileon density, i.e. Eq.~\eqref{Gal_dens_back}, or in terms of Jordan-frame quantities, the first line of Eq.~\eqref{effective_density}, will always be positive.

In this analysis we will restrict ourselves to models, where $\Omega_{\pi, \text{eff}}\coloneqq\rho_{\pi, \text{eff}}/\rho_{\text{crit}}$ is small and the background evolution is at least roughly inside of current observational limits. Since the main goal of this analysis is to get a qualitative understanding of the shape-dependence effect, we postpone a thorough data analysis to future works. 

We construct our background models with \texttt{hi\_class} in the following way: Our Galileon model has two free parameters $C_2$ and $C_3$. Since we are mostly interested in the effect of shape dependence which is proportional to $C_3$, we consider $C_2$ to be either zero or subdominant on the background level. Thus, the value of $\Omega_{\pi, \text{eff}, 0}$ depends mostly on $C_3$. When we give \texttt{hi\_class} a goal value for $\Omega_{\pi, \text{eff}, 0}$ as input, the code performs a shooting algorithm that fits the value of $C_3$ corresponding to the given $\Omega_{\pi, \text{eff}, 0}$. The parameter of $\Omega_\Lambda$ will always be used to fulfil the closure condition $1= \sum_i \Omega_{i,0}$. This also ensures that the sum of the energy densities driving the late time acceleration of the Universe $\Omega_{DE}\coloneqq \Omega_\Lambda+\Omega_{\pi, \text{eff}}$ will be identical to $\Omega_\Lambda$ in a purely $\Lambda$CDM model.

The Galileon models that we use throughout this work, labelled Gal 1-5, are defined in Table \ref{tab:models}. The models Gal 1-3 enable us to study the effects of increasing $\Omega_{\pi, \text{eff}, 0}$, whereas the models Gal 4 and Gal 5 test the influences of the parameter $C_2$ while keeping $\Omega_{\pi, \text{eff}, 0}$ constant. We compare these models to $\Lambda$CDM, which corresponds to vanishing $\Omega_{\pi, \text{eff}, 0}$ \jchanges{achieved by the limit $C_2\rightarrow 0$, $C_3\rightarrow\infty$ (compare Eq.~\eqref{pi_prime}).} \jchanges{Although \texttt{hi\_class} checks the stability of the considered models on the cosmological background, it is not guaranteed that these models will be stable on any background. However, for the sake of studying phenomenological apsects of Vainshtein screening, only stability on a cosmological background is paramount.}

\begin{table}
	\begin{center}
		\item[]\begin{tabular}{|l|l|l|l|}
			\hline
			Model name & $\Omega_{\pi, \text{eff}, 0}$ & $C_3\,[\SI{}{Mpc^2}]$ & $C_2$ \\
			\hline
			Gal 1 & $\SI{-0.01}{}$ & $\SI{4.276747e9}{}$ & $0$ \\
			Gal 2 & $\SI{-0.02}{}$ & $\SI{1.063205e9}{}$ & $0$ \\
			Gal 3 & $\SI{-0.03}{}$ & $\SI{4.699956e8}{}$ & $0$ \\
			Gal 4 & $\SI{-0.01}{}$ & $\SI{6.127518e9}{}$ & $\SI{-5.925926}{}$ \\
			Gal 5 & $\SI{-0.01}{}$ & $\SI{1.918030e9}{}$ & $\SI{5.925926}{}$ \\
			$\Lambda$CDM & $\SI{0}{}$ & $\SI{0}{}$ & $\SI{0}{}$ \\
			\hline
		\end{tabular}
	\end{center}
\caption{Definition of the Galileon models we use. The value of $C_3$ is obtained by means of a shooting algorithm in order to match the required $\Omega_{\pi, \text{eff}, 0}$. Gal 1, \dots, Gal 3 test the effects of $\Omega_{\pi, \text{eff}, 0}$, and Gal 4 and Gal 5 enable us to study the influences of $C_2$ while keeping $\Omega_{\pi, \text{eff}, 0}$ constant.}
\label{tab:models}
\end{table}

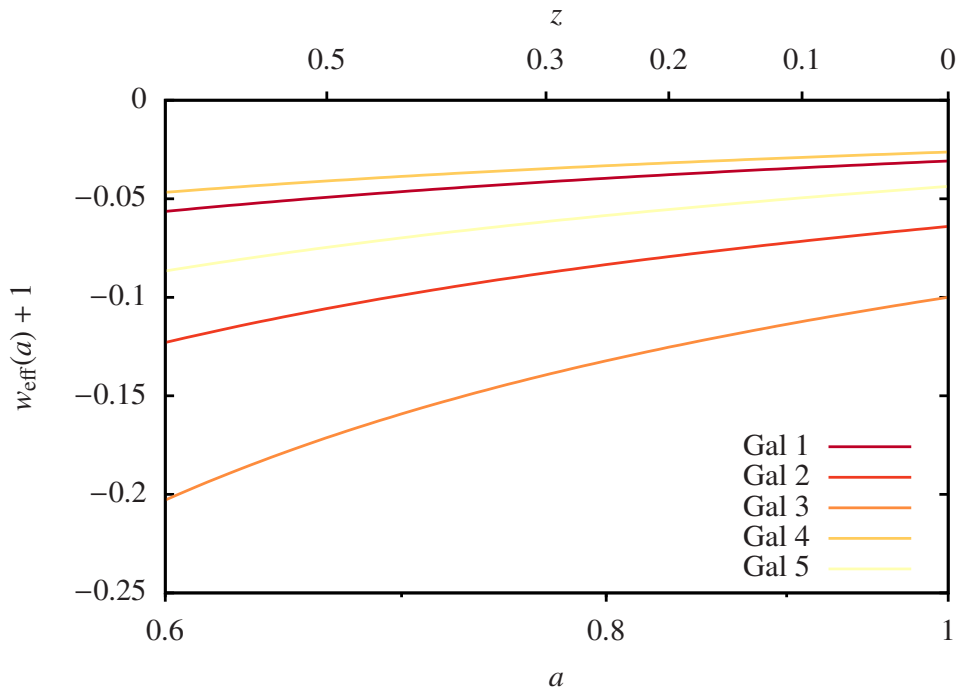
\begin{figure}
\begin{center}
	\item[]\resizebox{!}{0.6\textwidth}{\input{w_eff}}
\end{center}
\caption{Deviations from $-1$ of the effective equation of state parameter $w_{\text{eff}}$ of the energy components driving the late time acceleration of the Universe as a function of scale factor and redshift for the models Gal 1-5 defined in Table \ref{tab:models}. The dark energy density becomes dominant over the matter density at $z\approx 0.33$. }
\label{fig:w_eff}
\end{figure}

In order to check how much the inclusion of the Galileon field affects the background evolution, we consider an effective equation of state parameter $w_{\text{eff}}$ defined as:
\begin{eqnarray}
w_{\text{eff}} \coloneqq \frac{p_\Lambda+p_{\pi, \text{eff}}}{\rho_\Lambda+\rho_{\pi, \text{eff}}},
\end{eqnarray}
i.e.\ the equation of state parameter of all the energy components driving the late time acceleration of the Universe. The deviations of this quantity from $-1$ are plotted in Figure \ref{fig:w_eff}. The value of $w_{\text{eff}}$ is always smaller than $-1$ but tends towards $-1$ at late times. Deviations from $-1$ become large at early times, but dark energy only becomes dominant over the matter density at $z\approx 0.33$ for all considered models. Current bounds on the equation of state parameter of dark energy from the DES \cite{DES} indicate the values $w_p=-1.01^{+0.04}_{-0.04}$ and $w_a=-0.28^{+0.37}_{-0.48}$ for a parametrisation $w=w_p+w_a\,(a_p-a)$ with the pivot redshift $z_p=1/a_p-1=0.2$.

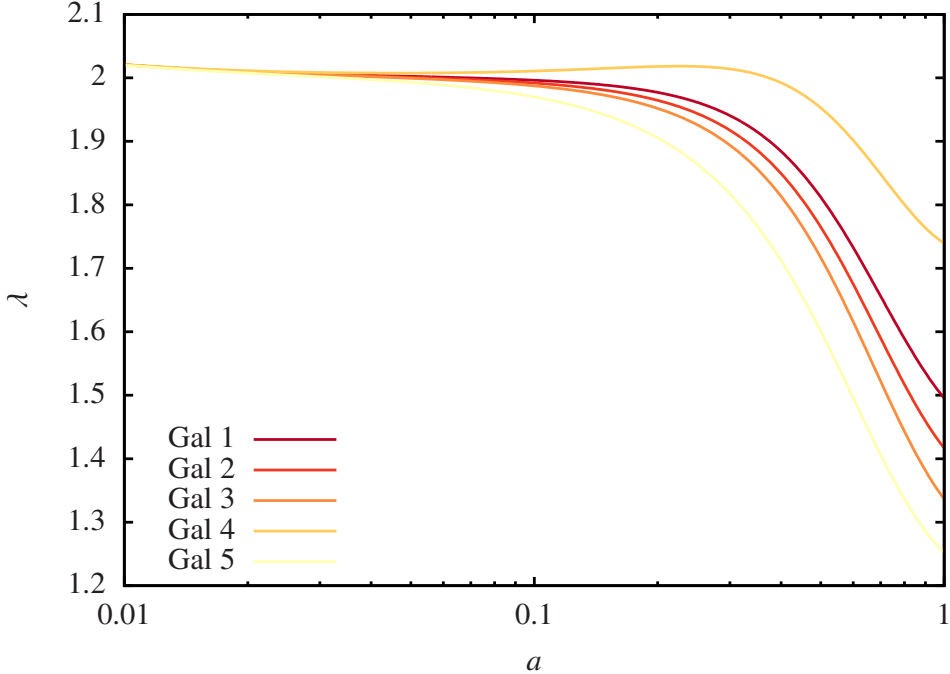
\begin{figure}
	\begin{center}
		\item[]\resizebox{!}{0.6\textwidth}{\input{perturbation}}
	\end{center}
	\caption{The quantity $\lambda$ defined in Eq.~\eqref{condition} \jchanges{describing} the relative size of the non-linearity scales of \jchanges{standard perturbation theory with respect to} the cubic Galileon is plotted for the models Gal 1-5 defined in Table \ref{tab:models}. If this quantity is larger than $1$, perturbativity of the cubic Galileon is assured as long as \jchanges{the standard condtion} $\delta^{(1)}\ll 1$ \jchanges{is fulfilled}. Although this is fulfilled for all the models studied in this work, it is not a trivial test and should be done for all future perturbative analyses of theories with Vainshtein screening.}
	\label{fig:perturbation}
\end{figure}

Before we compute the first and second order perturbations, we have to \jchanges{check the condition \eqref{condition} for perturbativity of the Galileon equation of motion} as outlined in Section \ref{sec:condition}. For this we plot the quantity $\lambda$ defined by Eq.~\eqref{condition} as a function of time in Figure \ref{fig:perturbation}. We find that the condition $\lambda\gtrsim1$ is satisfied for all the considered models and for the entire evolution of the Universe, confirming the validity of our perturbative analysis. \jchanges{Figure \ref{fig:perturbation} confirms our analytic prediction that $\lambda=2$ during matter domination. However, we also observe that $\lambda$ decreases over time in the late Universe, suggesting that perturbativity of the Galileon equation might be more restricted for a de-Sitter Universe. In particular, t}he value of $C_2$ appears to have significant impact on the late behaviour of $\lambda$. \jchanges{We leave a more detailed analysis of these phenomena for future work.}

\subsection{Linear growth}
\label{sec:numerics_linear}

In this section we compute the linear growth rate numerically. In order to obtain the Green's function in Eq.~\eqref{Greens_function}, we need two independent solutions of the linear growth equation \eqref{lin_pert} -- let them be $D_1$ and $D_2$. To obtain them, we solve Eq.~\eqref{lin_pert} for two different initial conditions. To establish an approximate connection between the solution $D_1$ and the growing mode and solution $D_2$ and the decaying mode, respectively, we set the initial conditions to be the solutions of the Meszaros equation \cite{Meszaros:1974tb,1975A&A....41..143G}, valid during radiation and matter domination:\jchanges{
\begin{eqnarray}
D_1(a_i) &=& 2+3y_i \nonumber \\
D_2(a_i) &=& (4+6y_i)\coth^{-1}\left( \sqrt{1+y_i} \right)-6\sqrt{1+y_i},
\end{eqnarray}}
where $y\coloneqq a/a_{eq}$ is the scale factor relative to the scale factor at radiation-matter equality $a_{eq}$. The \jchanges{thereby obtained solutions $D_1(a)$ and $D_2(a)$ are afterwards} normalized such that $D_1(a=1)=1$. 

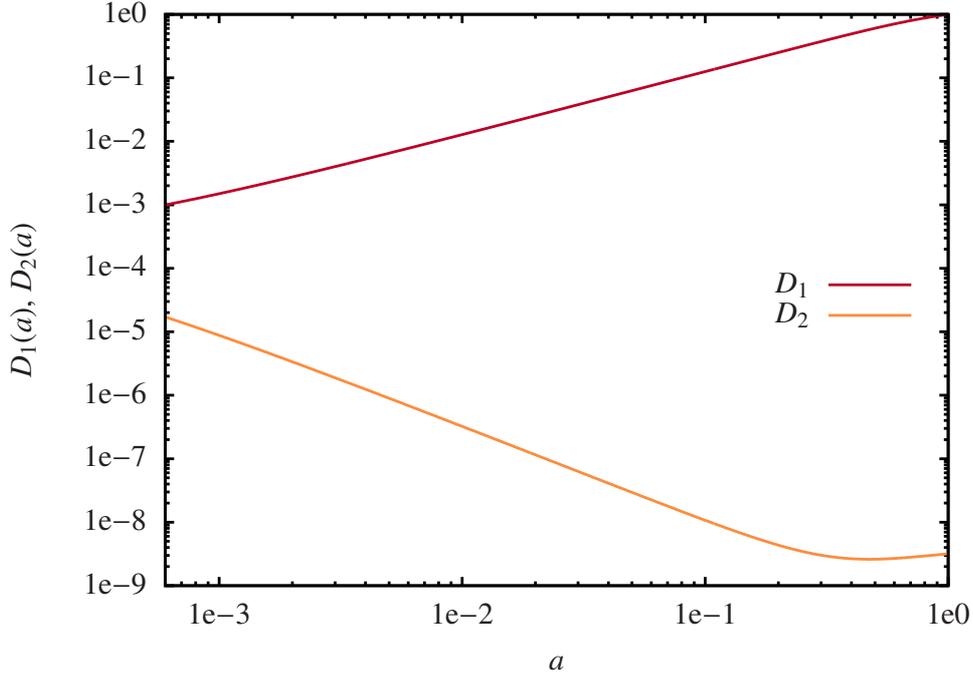
\begin{figure}
	\begin{center}
		\item[]\resizebox{!}{0.6\textwidth}{\input{growth_decay}}
	\end{center}
	\caption{The two independent solutions $D_1$ and $D_2$ for the model Gal 1 as a function of the scale factor. While the solution $D_1$ is equivalent to the growing mode, solution $D_2$ can roughly be associated with the decaying mode. The solutions are normalised such that $D_1(a=1)=1$.}
	\label{fig:growth_decay}
\end{figure}

The results of the integration for the model Gal 1 are shown in Figure \ref{fig:growth_decay}. Since we assume that the Universe is dominated by matter and radiation when setting the initial conditions, $D_2$ deviates slightly from the true decaying mode, showing a final residual of $\sim 10^{-9}D_1(a=1)$. This small deviation is not a concern for the purposes of this work as all we need is the growing mode and another independent solution. 

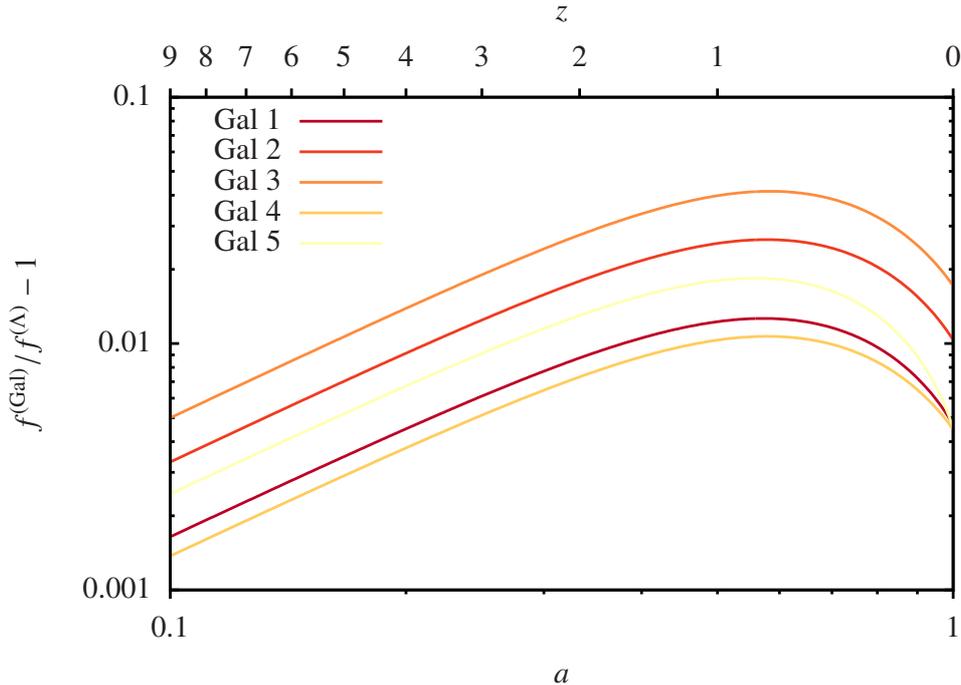
\begin{figure}
	\begin{center}
		\item[]\resizebox{!}{0.6\textwidth}{\input{growth_function}}
	\end{center}
	\caption{The deviations of the linear growth rate $f$ from the $\Lambda$CDM growth rate for the models Gal 1-5 defined in Table \ref{tab:models} as a function of scale factor and redshift.}
	\label{fig:growth_function}
\end{figure}

In order to quantify the linear growth in our models we consider the growth rate $f$. In Figure \ref{fig:growth_function} we present deviations of $f$ from the $\Lambda$CDM result $f_{\Lambda\text{CDM}}$. We see that deviations do not exceed $\SI{5}{\percent}$. This is roughly within current observational bounds, which indicate order $\SI{10}{\percent}$ relative uncertainties for $f\sigma_8$ assuming a $\Lambda$CDM cosmology, where $\sigma_8$ is the amplitude of the power spectrum. See Ref.~\cite{Nesseris:2017vor} for a compendium on past constraints on $f\sigma_8$ and Refs.~\cite{Zhao:2018jxv, Mohammad:2018mdy, Alam:2016hwk, Shi:2017qpr, Carrick:2015xza, Huterer:2016uyq} for some recent developments. 

\subsection{The matter bispectrum}
\label{sec:numerics_bispectrum}

We can now integrate the functions $\mathcal{B}_{GR}$ and $\mathcal{B}_\pi$ in Eq.~\eqref{curly_B} and Eq.~\eqref{curly_P}, where we recall that $\mathcal{B}_\pi$ and $\mathcal{B}_{GR}$ respectively describe the contributions to the bispectrum from the shape dependence of Vainshtein screening, and the non-linearities coming from the continuity and Euler equations.

\begin{figure}
	\begin{center}
		\item[]\resizebox{!}{0.6\textwidth}{\input{bispectrum}}
	\end{center}
	\caption{The relative difference between $\mathcal{B}=\mathcal{B}_\pi+\mathcal{B}_{GR}$ for the Galileon models Gal 1-5 defined in Table \ref{tab:models} and the $\Lambda$CDM model. The function $\mathcal{B}_\pi$ describes the effect of the shape-dependence on the form factor $F_2$, the function $\mathcal{B}_{GR}$ is a standard GR contribution of $F_2$ which is degenerate with $\mathcal{B}_\pi$, see Eq.~\eqref{F_2}. The slope of $a^{-3/2}$ of $\mathcal{B}_\pi$ during matter domination is a distinctive prediction of our model.}
	\label{fig:bispectrum}
\end{figure}
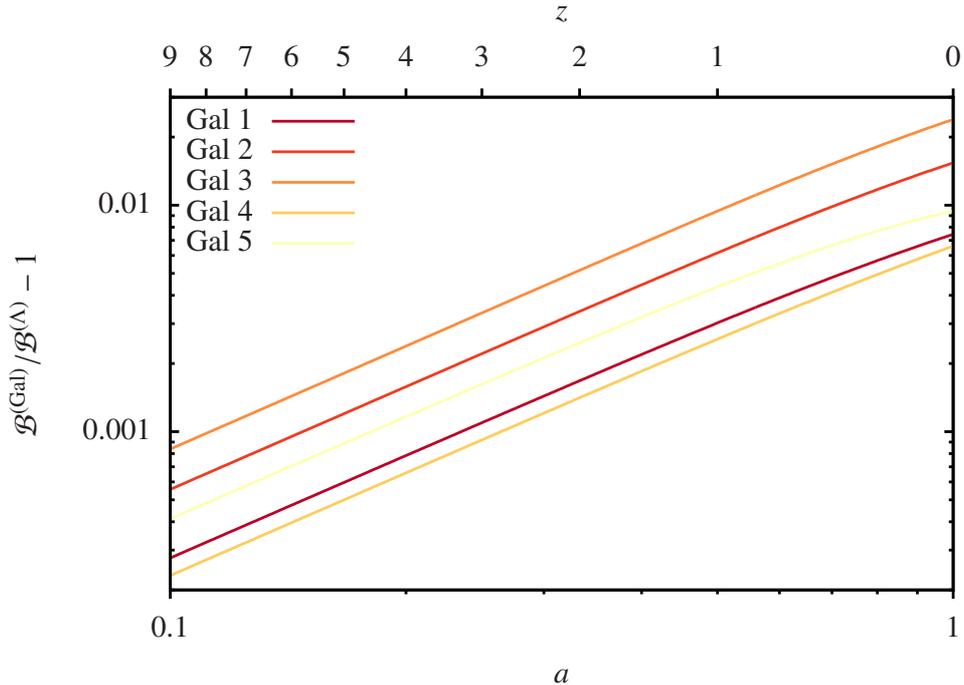

In Figure \ref{fig:bispectrum}, we show the relative difference between the sum of both contributions $\mathcal{B}=\mathcal{B}_\pi+\mathcal{B}_{GR}$ for the five Galileon models and $\Lambda$CDM. The relative difference scales as $\Omega_\pi\propto a^{3/2}$ during matter domination, in agreement with our analytical result derived in Eq.~\eqref{analytic}. The slope remains approximately the same at late times, but it appears to be sensitive to the value of $C_2$. In fact, we can see that the Galileon model Gal 5, characterised by $C_2 > 0$, displays a much shallower slope at $z<1$ compared to the other models.

In general, deviations in the bispectrum from $\Lambda$CDM are larger for models which also display significant modifications in the background evolution and linear growth rate, as illustrated by Figures \ref{fig:w_eff} and \ref{fig:growth_function}. We observe the largest deviations in the bispectrum for the model Gal 3, where they are of order $2-3\SI{}{\percent}$ at redshift $z=0$.

The deviations, with respect to $\Lambda$CDM, of the term $\mathcal{B}$ in the Galileon models have two different origins. First, the term $\mathcal{B}_\pi$ is altogether absent in $\Lambda$CDM, and second, the term $\mathcal{B}_{GR}$ depends on the evolution of linear perturbations (see Eq.~\eqref{curly_B} and Figure \ref{fig:growth_function}), which is also modified for the Galileons.
 
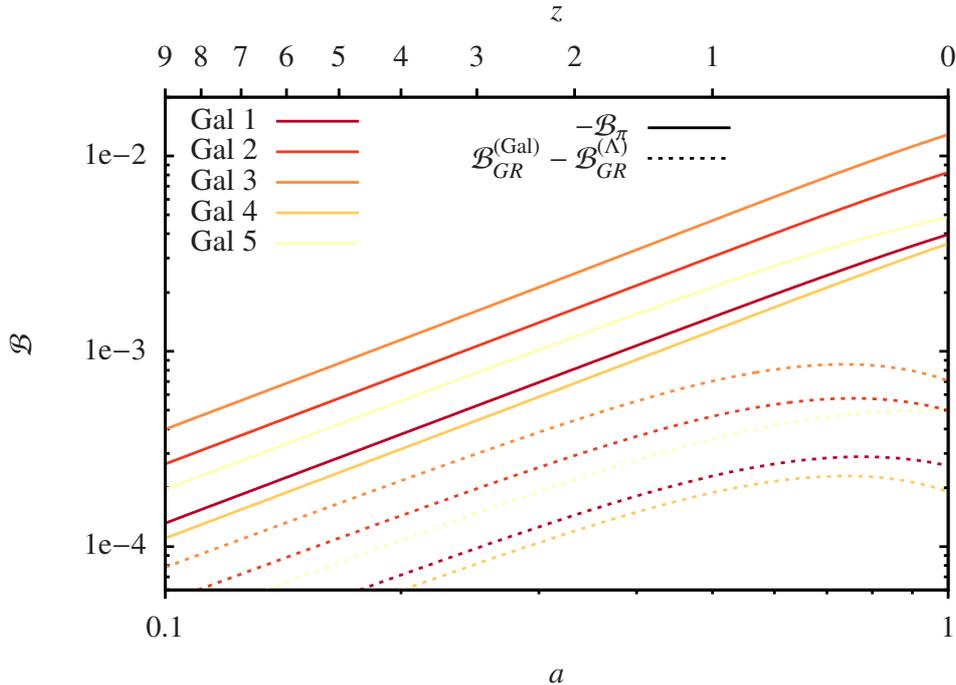
\begin{figure}
	\begin{center}
		\item[]\resizebox{!}{0.6\textwidth}{\input{curly_B}}
	\end{center}
	\caption{Comparison of $\mathcal{B}_\pi$ (solid lines) and $\mathcal{B}_{GR}^{\rm (Gal)}-\mathcal{B}_{GR}^{(\Lambda)}$ (dashed lines), for the Galileon models Gal 1-5 defined in Table \ref{tab:models}. While $\mathcal{B}_\pi$ represents the effect of the shape-dependence on the bispectrum, $\mathcal{B}_{GR}^{\rm (Gal)}-\mathcal{B}_{GR}^{(\Lambda)}$ measures the modification of the bispectrum due to the altered evolution of linear perturbations. $\mathcal{B}_{GR}^{\rm (Gal)}$ and $\mathcal{B}_{GR}^{(\Lambda)}$ describe a GR contribution evaluated, respectively, on a Galileon or a $\Lambda$CDM background.}
	\label{fig:curly_B}
\end{figure}

We compare these two contributions in Figure \ref{fig:curly_B}, to determine which is dominant. For the linear-evolution term, we display the difference between $\mathcal{B}_{GR}^{\rm (Gal)}$ and $\mathcal{B}_{GR}^{(\Lambda)}$, i.e. the term $\mathcal{B}_{GR}$ evaluated on a Galileon or a standard $\Lambda$CDM background. We can see that the shape-dependence in the Galileon equation of motion \eqref{shape_term} is the dominant effect modifying the bispectrum compared to a $\Lambda$CDM cosmology.

The reason why $\mathcal{B}_{GR}^{\rm (Gal)}-\mathcal{B}_{GR}^{(\Lambda)}$ is subdominant can be understood from the definition of $\mathcal{B}_{GR}$, and generalises to other non-$\Lambda$CDM cosmologies. Eq.~\eqref{curly_B} shows that $\mathcal{B}_{GR}(\tau)$ is an integral over the time derivative of $D_+$, normalized by $D_+(\tau)$. This is quite intuitive since $\mathcal{B}_{GR}$ describes the non-linearities in the continuity and Euler equations due to the total time derivative $\mathrm{d}/\mathrm{d}t=\partial/\partial t+\vec{v}\cdot\nabla$ depending on the flow of matter. Since the first-order continuity equation in the quasi-static approximation reads $\Delta v^{(1)}=-\delta^{(1)\prime}$, velocities are related to the time derivative of the density fluctuations. 

Since $D_+'$ is normalized by $D_+(\tau)$ in the definition of $\mathcal{B}_{GR}(\tau)$ independent of the underlying cosmology, the difference between $D_+^{\rm (Gal)}$ and $D_{+}^{(\Lambda)}$ must be either growing at early times and decaying at late times or vice versa. This means that 
\begin{equation}
\frac{D_+^{\prime\,\rm (Gal)}(\tilde{\tau})}{D_+^{\rm (Gal)}(\tau)}-\frac{D^{\prime\,(\Lambda)}_{+}(\tilde{\tau})}{D_{+}^{(\Lambda)}(\tau)}
\end{equation}
must change its sign at some time $\tilde{\tau}<\tau$. Consequently, $\mathcal{B}_{GR}^{\rm (Gal)}-\mathcal{B}_{GR}^{(\Lambda)}$ will always be small since negative and positive contributions cancel out when integrating. We conclude from this that, quite generally, intrinsically non-linear modifications of gravity, like the shape-dependent term in the Galileon equation of motion Eq.~\eqref{shape_term}, will have a larger impact on the bispectrum than linear modifications which influence the bispectrum indirectly through $\mathcal{B}_{GR}$.

\begin{figure}
	\begin{center}
		\item[]\resizebox{!}{0.6\textwidth}{\input{red_bispec}}
	\end{center}
	\caption{The relative difference of the reduced bispectrum for the Galileon models Gal 1-5 defined in Table \ref{tab:models} with respect to the $\Lambda$CDM model at $z=0$ is plotted against $\mu_{12}$ -- the cosine of the angle between $\vec{k}_1$ and $\vec{k}_2$. In plot $(a)$ the wavenumbers $k_1$ and $k_2$ are equal and in plot $(b)$ $k_1=2\times k_2$; in both cases $k_1=\SI{0.1}{Mpc/h}$.}
	\label{fig:red_bispec}
\end{figure}
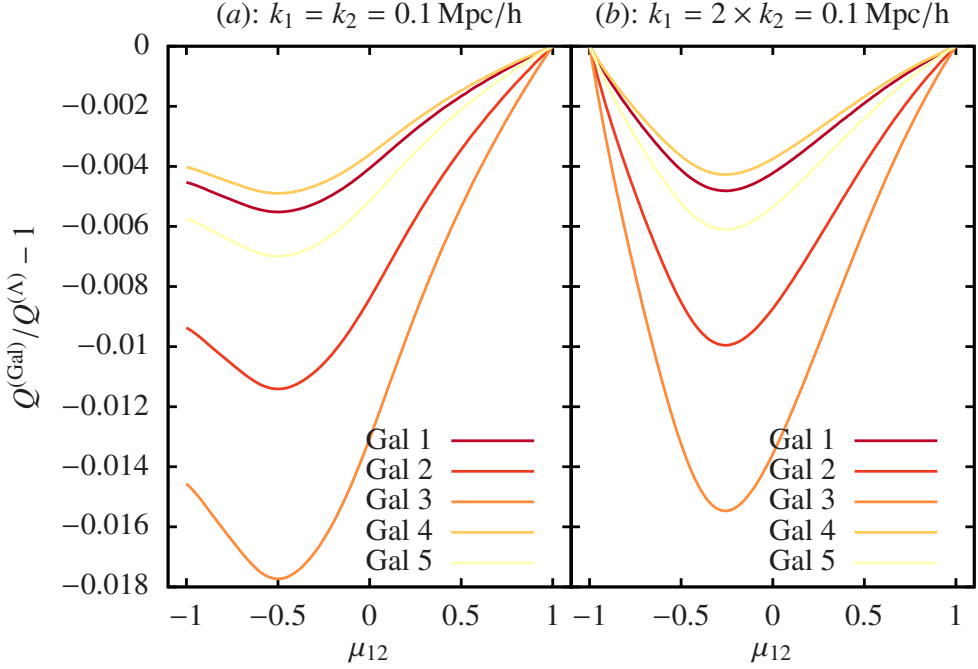

Finally, we consider the reduced bispectrum in Eq.~\eqref{red_bispec}, which is mostly independent of scale and linear growth. In Figure \ref{fig:red_bispec}, we show the relative difference between the reduced bispectrum for the Galileon models and $\Lambda$CDM. The triangle $\vec{k}_1+\vec{k}_2+\vec{k}_3=0$ is parametrized by $\mu_{12}$, the cosine of the angle between $\vec{k}_1$ and $\vec{k}_2$, and the absolute values of $\vec{k}_1$ and $\vec{k}_2$. In both plots in Figure \ref{fig:red_bispec} we keep $k_1$ and $k_2$ constant while varying $\mu_{12}$; in the left panel ($a$), we set $k_1=k_2$, whereas in the right panel ($b$), we use $k_1=2\times k_2$.

In panel $(a)$, deviations from $\Lambda$CDM vanish for $\mu_{12}=1$. This corresponds to the flattened limit of the triangle with $k_3=2\times k_1$, shown in the right panel. This is in agreement with our expectations that planar symmetry, corresponding to the flattened limit, will result in no Vainshtein screening -- see Section \ref{sec:second_order}. The deviations from $\Lambda$CDM are maximal for $\mu_{12}=-0.5$, which corresponds to an equilateral triangle, i.e.\ the most symmetric configuration. The signal is only slightly smaller in the squeezed limit $\mu_{12}=-1$.
In plot $(b)$, where $k_1=2\times k_2$, $\mu_{12}=-1$ and $\mu_{12}=1$ both correspond to the flattened limit and the signal vanishes. The signal is maximal for $\mu=-0.25$ which corresponds to $k_3=k_1$, i.e.\ an isosceles triangle, which, for $k_1=2\times k_2$ is the most symmetric configuration. 
In Refs. \cite{Japanese, Bellini2015} a qualitatively similar behaviour of the reduced bispectrum in general Horndeski theories was observed, but without making the connection to the shape dependence of Vainshtein screening.

Summarizing, the shape dependence of Vainshtein screening as seen by the bispectrum is perfectly analogous to the shape dependence in real space outlined in Section \ref{Shape_dependence}: in real space, the non-linearities responsible for Vainshtein screening are larger the more evenly the field depends on all three directions of space; in the bispectrum, the non-linearities are largest for the most symmetric triangle configurations.

\section{Summary and Conclusion}
\label{Conclusion}

In this work, we investigated the effects of the shape dependence of Vainshtein screening on the cosmic matter bispectrum. The dependence of Vainshtein screening on the shape of the source mass was first found in Ref.~\cite{Clare_shape_dependence}, which showed that more symmetric sources led to a more effective suppression of the fifth force. If Vainshtein screening is present in nature, we can then expect that it will leave an imprint on cosmic structures, given that the cosmic web is characterised by many differently shaped objects like walls, filaments and clusters. 

The simplest modified gravity model displaying the Vainshtein screening mechanism is the cubic Galileon, which we used as a proxy to test the effects of shape-dependence on the matter bispectrum. Unlike previous work on the bispectrum in Galileon theories, we assumed the Galileon field to be conformally coupled to matter, so as to make the fifth force mediated by the Galileon field explicit. We also restricted ourselves to models where the Galileon energy density is subdominant throughout the entire evolution of the Universe, to reflect constraints ruling out the Galileon as the single component driving the accelerated expansion of the Universe.

Previous analyses \cite{Matarrese, Japanese, Bellini2015} seemed to indicate that there was no qualitatively new information in the bispectrum that was not already present in the power spectrum, which is also easier to measure. However, in this work we have shown that a unique signature emerges in the bispectrum, imprinted by the shape dependence of Vainshtein screening, that would not be observable in the power spectrum alone. We performed an analytic, perturbative analysis of our coupled Galileon model in the Einstein frame, checking explicitly that the emergence of a new non-linearity scale associated to the Vainshtein radius does not lead to a breakdown of perturbation theory.

The non-linearities of Vainshtein screening leave an imprint on the form factor $F_2(\vec{k}_i,\vec{k}_j)$ of the reduced bispectrum, adding a time-dependent term $\mathcal{B}_\pi(\tau)(1-\mu^2)$ varying with the cosine $\mu$ of the angle between two wavevectors $\vec{k}_i,\vec{k}_j$. This $\mu$ dependence reflects the shape dependence found in \cite{Clare_shape_dependence}: in fact, we observe that the non-linearities are largest for the most symmetric configuration of the bispectrum triangle -- i.e. the equilateral one -- whereas they vanish for the flattened limit, which corresponds to planar symmetry in real space.

A simplified, analytic computation of $\mathcal{B}_\pi$ assuming matter domination revealed that $\mathcal{B}_\pi$ scales with the Galileon density, which was confirmed by numerical analysis using the \texttt{hi\_class} code. The effect of the shape dependence on the matter bispectrum is found to be at percent level today for Galileon models where the Galileon energy density $\Omega_{\pi, \text{eff}, 0}$ is at percent level as well.

Critically, the shape dependence of Vainshtein screening displays a distinctive time dependence $\propto a^{3/2}$, which is dominant over corrections originating from modifications of the linear growth. This signature is an independent observational effect from modifications of the background evolution or the linear growth factor, which can be similar for different models of modified gravity. As such, it may be used to break degeneracies between theories of modified gravity predicting similar deviations in both.

\acknowledgments
We acknowledge use of the \texttt{hi\_class} code \jchangesnew{\cite{Bellini:2019syt, hi_class, Blas:2011rf}}. The authors thank Miguel Zumalacarregui, Emilio Bellini and Ignacy Sawicki for generously sharing \jchangesnew{the latest} version of \texttt{hi\_class} that can evolve the exact Galileon equation of motion numerically \jchangesnew{prior to its official release}. The authors also thank David Parkinson and David Seery for helpful discussions. This work was supported by a Research Leadership Award from the Leverhulme Trust. CB is also supported by a Royal Society University Research fellowship.

\appendix

\section{Formulas and Definitions}
\label{appendix}

The Conservation equations, Einstein equations and the Galileon field equation at linear order:
\begin{eqnarray}
\delta^{(1)\prime}&=&-\Delta v^{(1)}+3\phi^{(1)\prime}+\frac{1}{2(1+\pi)}\left( \pi^{(1)\prime}-\pi'\frac{\pi^{(1)}}{1+\pi} \right), \label{Continuity} \\
v^{(1)\prime}+\mathcal{H}v^{(1)}&=&-\psi^{(1)}-\frac{1}{2(1+\pi)}\left( \pi'v^{(1)}+\pi^{(1)} \right), \label{Euler} \\
-\frac{a^2G^0_0}{M^2_p}&=& 2\Delta\phi^{(1)}=3a^2\bar{\rho}\delta^{(1)}+\frac{C_3\pi'^2}{a^2}\Delta\pi^{(1)}, \label{Poisson} \\
a^2\partial_i^{-1}\partial_j^{-1}G^{i}_j&=& \phi^{(1)}-\psi^{(1)}=0, \label{gravitational_slip}
\end{eqnarray}
\begin{equation}
\left(C_2-\frac{2C_3}{a^2}(\pi''+\mathcal{H}\pi')\right)\Delta\pi^{(1)}-\frac{C_3\pi'^2}{a^2}\Delta\phi^{(1)}=\frac{3\bar{\rho}a^2}{2(1+\pi)}\left( \delta^{(1)}-\frac{\pi^{(1)}}{1+\pi} \right). \label{Galileon_pre}
\end{equation}
The functions $\mathcal{A}_{GR}$, $\mathcal{B}_{GR}$, $\mathcal{B}_\pi$ used to describe $F_2$ are defined by:
\begin{eqnarray}
\mathcal{A}_{GR}(\tau)&\coloneqq& 2\int_{\tau_i}^{\tau}\mathrm{d}\tilde{\tau}\,G(\tau, \tilde{\tau}
)\left( f^2(\tilde{\tau})\mathcal{H}^2(\tilde{\tau}) +\alpha_\delta(\tilde{\tau}) \right)\frac{D_+^2(\tilde{\tau})}{D_+^2(\tau)}, \\
\mathcal{B}_{GR}(\tau)&\coloneqq& 2\int_{\tau_i}^{\tau}\mathrm{d}\tilde{\tau}\,G(\tau, \tilde{\tau}
)f^2(\tilde{\tau})\mathcal{H}^2(\tilde{\tau}) \frac{D_+^2(\tilde{\tau})}{D_+^2(\tau)} \nonumber \\
&=&\frac{2}{D_+^2(\tau)}\int_{\tau_i}^\tau \mathrm{d}\tilde{\tau}\, G(\tau, \tilde{\tau})\left(D_+'\right)^2, \label{curly_B} \\
\mathcal{B}_\pi(\tau)&\coloneqq& -2\int_{\tau_i}^{\tau}\mathrm{d}\tilde{\tau}\,G(\tau, \tilde{\tau}
)\frac{9C_3\bar{\rho}^2a^2\alpha_\pi^3}{8A^3(\tau)}\frac{D_+^2(\tilde{\tau})}{D_+^2(\tau)}. \label{curly_P} 
\end{eqnarray}
With the transformation Eq.~\eqref{transformation}, the action Eq.~\eqref{action} can brought into the form of a Horndeski theory defined by:
\begin{equation}
S[g_{\mu\nu}, \pi]=\int\mathrm{d}^4x\sqrt{-g}\left[ \sum_{i=2}^5\mathcal{L}_i +\mathcal{L}_m[g_{\mu\nu}, \psi_M] \right], \label{Horndeski}
\end{equation}
with
\begin{eqnarray}
\mathcal{L}_2&\coloneqq& K(\pi, X) \\
\mathcal{L}_3&\coloneqq& -G_3(\pi, X)\Box\pi \\
\mathcal{L}_4&\coloneqq& G_4(\pi, X)R +G_{4, X}\left[ (\Box\pi)^2 -(\nabla_\mu\nabla_\nu\pi)^2 \right] \\
\mathcal{L}_5&\coloneqq& G_5(\pi, X)G_{\mu\nu}\nabla^\mu\nabla^\nu\pi-\frac{1}{6}G_{5, X}(\pi, X)\left[ (\Box\pi)^3-3(\Box\pi)(\nabla_\mu\nabla_\nu\pi)^2 \right. \nonumber \\
& &\left.+2\nabla^\mu\nabla_\alpha\pi\nabla^\alpha\nabla_\beta\pi\nabla^\beta\nabla_\mu\pi \right],
\end{eqnarray}
where $2X\coloneqq -(\nabla\pi)^2$, $G_{\mu\nu}$ is the Einstein tensor and the $G_i$ are arbitrary functions of $\pi$ and $X$. The Horndeski functions for our model are given by:
\begin{eqnarray}
K(\pi, X)&=& - \frac{\Lambda_c}{(1+\pi)^2} -\frac{1}{1+\pi}\left( -C_2+\frac{3}{2(1+\pi)^2} \right)X \nonumber \\
& &-\frac{2C_3}{(1+\pi)}X^2, \label{K} \\
G_3(\pi, X)&=&-C_3 X , \label{G_3} \\
G_4(\pi, X)&=&\frac{1}{2(1+\pi)}. \label{G_4} \\
G_5(\pi, X)&=& 0.
\end{eqnarray}
Transformation rules between Einstein and Jordan frame (Jordan-frame quantities are marked by a tilde):
\begin{eqnarray}
\tilde{a}&=& \sqrt{1+\pi}a, \\
\tilde{\mathcal{H}}&=& \mathcal{H}+\frac{\pi'}{2(1+\pi)},  \\
\tilde{\rho}_m &=&\left( 1+\pi \right)^{-2}\rho_m.
\end{eqnarray}

\section{\boldmath The source term $S^{(\delta)}$}
\label{sec:source_term}

Here we give the full expression for the source term $S^{(\delta)}$ in terms of the source terms $S^{(1)}$, $S^{(4)}$, $S^{(5)}$, $S^{(6)}$ and $S^{(7)}$ which can be found in the appendix of \cite{Matarrese}. Since we have a Galileon model with a conformal coupling some of the source terms computed in \cite{Matarrese} have to be expanded for our model:
\begin{eqnarray}
\tilde{S}^{(5)}&\coloneqq& S^{(5)}+\frac{\bar{\rho}}{(1+\pi)^2}\pi^{(1)}\delta^{(1)}, \label{modified_S5} \\
\tilde{S}^{(6)}&\coloneqq& S^{(6)}+\frac{1}{1+ \pi}\left( \delta^{(1)}\pi^{(1)\prime}-\frac{\pi'\delta^{(1)}\pi^{(1)}+\pi^{(1)}\pi^{(1)\prime}}{1+\pi} \right), \label{modified_S6} \\
\tilde{S}^{(7)}&\coloneqq& S^{(7)}+\frac{1}{(1+\pi)^2}\left( \left(\partial_i\pi^{(1)}\right) \left( \partial^i\pi^{(1)} \right) +\pi^{(1)}\Delta\pi^{(1)} \right) \nonumber \\
& &-\frac{1}{1+\pi}\left( \left(\partial_i\delta^{(1)}\right)\left(\partial^i\pi^{(1)}\right) +\delta^{(1)}\Delta\pi^{(1)} \right) \nonumber \\
& &-\frac{\pi'}{1+\pi}\left( \partial_i\delta^{(1)}\partial^{i}v^{(1)}+ \delta^{(1)}\Delta v^{(1)} \right). \label{modified_S7}
\end{eqnarray}
The source term $S^{(\delta)}$ can now be defined as: 
\begin{eqnarray}
S^{(\delta)}&=&-\left( 1+\frac{C_3\pi'^2\alpha_\pi}{2a^2A(\tau)} \right)\left( \frac{S^{(1)}}{2}-\frac{S^{(4)}}{k^2} \right) +\frac{\alpha_\pi}{2A(\tau)}\tilde{S}^{(5)} \nonumber \\
& &+\tilde{S}^{(6)\prime}+\tilde{S}^{(6)}\left( \mathcal{H}+\frac{\pi'}{2(1+\pi)} \right)-\tilde{S}^{(7)}. \label{source_term}
\end{eqnarray}

\section{\boldmath Simplification of the form factor $F_2$}
\label{sec:F_2_simplification}

In this section we want to prove that $\mathcal{A}_{GR}+\mathcal{B}_{GR}=2$ which greatly simplifies the form factor $F_2$ in Eq.~\eqref{F2_extended}. For this, we firstly note that the following differential equation holds for the Wronskian $W$ defined in Eq.~\eqref{Wronskian}:
\begin{eqnarray}
W'= -\left( \mathcal{H}+\frac{\pi'}{2(1+\pi)} \right)W. \label{Wronskian_differential}
\end{eqnarray}
Now we can compute $\mathcal{A}_{GR}+\mathcal{B}_{GR}$. For simplicity of notation we will not write all of the dependencies on the integration variable $\tilde{\tau}$ explicitly, however, we will denote dependencies if they differ from $\tilde{\tau}$ or are crucial for the understanding of the equations.
\begin{eqnarray}
\mathcal{A}_{GR}+\mathcal{B}_{GR}&=& 2\int_{\tau_i}^\tau\mathrm{d}\tilde{\tau}\,G(\tau, \tilde{\tau} )\left( 2f^2\mathcal{H}^2 + \alpha_\delta \right)\frac{D_+^2(\tilde{\tau})}{D_+(\tau)} \label{step1} \\
&=& \frac{2}{D_+^2(\tau)}\int_{\tau_i}^\tau\mathrm{d}\tilde{\tau}\,W^{-1}(\tilde{\tau})\left( D_-(\tau)D_+(\tilde{\tau})- D_+(\tau)D_-(\tilde{\tau}) \right) \nonumber \\
& &\qquad \qquad\times\left( 2 D_+^{\prime 2} +D_+^{\prime\prime}D_+ +\left( \mathcal{H}+\frac{\pi'}{2(1+\pi)} \right)D_+'D_+ \right) \label{step2}
\end{eqnarray}
From Eq.~\eqref{step1} to Eq.~\eqref{step2} we used that the linear growth equation Eq.~\eqref{lin_pert} holds for the growth function $D_+$. Now we will integrate parts of this integral by parts:
\begin{eqnarray}
\mathcal{A}_{GR}+\mathcal{B}_{GR} &\ni& \frac{2}{D_+^2(\tau)}\int_{\tau_i}^\tau\mathrm{d}\tilde{\tau}\frac{D_+^{\prime\prime}D_+}{W}\left( D_-(\tau)D_+(\tilde{\tau})- D_+(\tau)D_-(\tilde{\tau}) \right) \nonumber \\
&=& -\frac{2}{D_+^2(\tau)}\int_{\tau_i}^\tau\mathrm{d}\tilde{\tau}\, D_+'\left[\left( \frac{D_+'}{W} -\frac{W'D_+}{W^2} \right)\left( D_-(\tau)D_+(\tilde{\tau})- D_+(\tau)D_-(\tilde{\tau}) \right)\right. \nonumber \\
& &\qquad\qquad\qquad\qquad\left.+ \frac{D_+}{W}\left( D_-(\tau)D_+'(\tilde{\tau})- D_+(\tau)D_-'(\tilde{\tau}) \right) \right].
\end{eqnarray}
Substituting this result back into the full expression Eq.~\eqref{step2} for $\mathcal{A}_{GR}+\mathcal{B}_{GR}$ and using the equation Eq.~\eqref{Wronskian_differential}, we arrive at:
\begin{eqnarray}
\mathcal{A}_{GR}+\mathcal{B}_{GR} &=& \frac{2}{D_+^2(\tau)}\int_{\tau_i}^\tau\mathrm{d}\tilde{\tau}\,W^{-1} \left[ D_+^{\prime 2}(\tilde{\tau})\left( D_-(\tau)D_+(\tilde{\tau})- D_+(\tau)D_-(\tilde{\tau}) \right)\right. \nonumber \\
& & \qquad\qquad\qquad\quad \left.- D_+(\tilde{\tau})D_+'(\tilde{\tau})\left( D_-(\tau)D_+'(\tilde{\tau})- D_+(\tau)D_-'(\tilde{\tau}) \right) \right] \nonumber \\
&=&\frac{2}{D_+(\tau)}\int_{\tau_i}^\tau\mathrm{d}\tilde{\tau}\,D_+'(\tilde{\tau})\frac{D_-'(\tilde{\tau})D_+(\tilde{\tau})-D_+'(\tilde{\tau})D_-(\tilde{\tau})}{W(\tilde{\tau})} \nonumber \\
&=& \frac{2}{D_+(\tau)}\int_{\tau_i}^\tau\mathrm{d}\tilde{\tau}\,D_+'(\tilde{\tau}) = 2.
\end{eqnarray}


\bibliographystyle{JHEP}
\bibliography{galileons}

\end{document}

%% file: w_eff.tex
\begingroup
  \makeatletter
  \providecommand\color[2][]{%
    \GenericError{(gnuplot) \space\space\space\@spaces}{%
      Package color not loaded in conjunction with
      terminal option `colourtext'%
    }{See the gnuplot documentation for explanation.%
    }{Either use 'blacktext' in gnuplot or load the package
      color.sty in LaTeX.}%
    \renewcommand\color[2][]{}%
  }%
  \providecommand\includegraphics[2][]{%
    \GenericError{(gnuplot) \space\space\space\@spaces}{%
      Package graphicx or graphics not loaded%
    }{See the gnuplot documentation for explanation.%
    }{The gnuplot epslatex terminal needs graphicx.sty or graphics.sty.}%
    \renewcommand\includegraphics[2][]{}%
  }%
  \providecommand\rotatebox[2]{#2}%
  \@ifundefined{ifGPcolor}{%
    \newif\ifGPcolor
    \GPcolortrue
  }{}%
  \@ifundefined{ifGPblacktext}{%
    \newif\ifGPblacktext
    \GPblacktexttrue
  }{}%
  \let\gplgaddtomacro\g@addto@macro
  \gdef\gplbacktext{}%
  \gdef\gplfronttext{}%
  \makeatother
  \ifGPblacktext
    \def\colorrgb#1{}%
    \def\colorgray#1{}%
  \else
    \ifGPcolor
      \def\colorrgb#1{\color[rgb]{#1}}%
      \def\colorgray#1{\color[gray]{#1}}%
      \expandafter\def\csname LTw\endcsname{\color{white}}%
      \expandafter\def\csname LTb\endcsname{\color{black}}%
      \expandafter\def\csname LTa\endcsname{\color{black}}%
      \expandafter\def\csname LT0\endcsname{\color[rgb]{1,0,0}}%
      \expandafter\def\csname LT1\endcsname{\color[rgb]{0,1,0}}%
      \expandafter\def\csname LT2\endcsname{\color[rgb]{0,0,1}}%
      \expandafter\def\csname LT3\endcsname{\color[rgb]{1,0,1}}%
      \expandafter\def\csname LT4\endcsname{\color[rgb]{0,1,1}}%
      \expandafter\def\csname LT5\endcsname{\color[rgb]{1,1,0}}%
      \expandafter\def\csname LT6\endcsname{\color[rgb]{0,0,0}}%
      \expandafter\def\csname LT7\endcsname{\color[rgb]{1,0.3,0}}%
      \expandafter\def\csname LT8\endcsname{\color[rgb]{0.5,0.5,0.5}}%
    \else
      \def\colorrgb#1{\color{black}}%
      \def\colorgray#1{\color[gray]{#1}}%
      \expandafter\def\csname LTw\endcsname{\color{white}}%
      \expandafter\def\csname LTb\endcsname{\color{black}}%
      \expandafter\def\csname LTa\endcsname{\color{black}}%
      \expandafter\def\csname LT0\endcsname{\color{black}}%
      \expandafter\def\csname LT1\endcsname{\color{black}}%
      \expandafter\def\csname LT2\endcsname{\color{black}}%
      \expandafter\def\csname LT3\endcsname{\color{black}}%
      \expandafter\def\csname LT4\endcsname{\color{black}}%
      \expandafter\def\csname LT5\endcsname{\color{black}}%
      \expandafter\def\csname LT6\endcsname{\color{black}}%
      \expandafter\def\csname LT7\endcsname{\color{black}}%
      \expandafter\def\csname LT8\endcsname{\color{black}}%
    \fi
  \fi
    \setlength{\unitlength}{0.0500bp}%
    \ifx\gptboxheight\undefined%
      \newlength{\gptboxheight}%
      \newlength{\gptboxwidth}%
      \newsavebox{\gptboxtext}%
    \fi%
    \setlength{\fboxrule}{0.5pt}%
    \setlength{\fboxsep}{1pt}%
\begin{picture}(7200.00,5040.00)%
    \gplgaddtomacro\gplbacktext{%
      \csname LTb\endcsname
      \put(1078,767){\makebox(0,0)[r]{\strut{}$-0.25$}}%
      \put(1078,1477){\makebox(0,0)[r]{\strut{}$-0.2$}}%
      \put(1078,2187){\makebox(0,0)[r]{\strut{}$-0.15$}}%
      \put(1078,2896){\makebox(0,0)[r]{\strut{}$-0.1$}}%
      \put(1078,3606){\makebox(0,0)[r]{\strut{}$-0.05$}}%
      \put(1078,4316){\makebox(0,0)[r]{\strut{}$0$}}%
      \put(1210,484){\makebox(0,0){\strut{}$0.6$}}%
      \put(4360,484){\makebox(0,0){\strut{}$0.8$}}%
      \put(6803,484){\makebox(0,0){\strut{}$1$}}%
      \put(2364,4599){\makebox(0,0){\strut{}0.5}}%
      \put(3930,4599){\makebox(0,0){\strut{}0.3}}%
      \put(4807,4599){\makebox(0,0){\strut{}0.2}}%
      \put(5759,4599){\makebox(0,0){\strut{}0.1}}%
      \put(6803,4599){\makebox(0,0){\strut{}$0$}}%
    }%
    \gplgaddtomacro\gplfronttext{%
      \csname LTb\endcsname
      \put(198,2541){\rotatebox{-270}{\makebox(0,0){\strut{}$w_{\text{eff}}(a)+1$}}}%
      \put(4006,154){\makebox(0,0){\strut{}$a$}}%
      \put(4006,4929){\makebox(0,0){\strut{}$z$}}%
      \csname LTb\endcsname
      \put(5816,1820){\makebox(0,0)[r]{\strut{}Gal 1}}%
      \csname LTb\endcsname
      \put(5816,1600){\makebox(0,0)[r]{\strut{}Gal 2}}%
      \csname LTb\endcsname
      \put(5816,1380){\makebox(0,0)[r]{\strut{}Gal 3}}%
      \csname LTb\endcsname
      \put(5816,1160){\makebox(0,0)[r]{\strut{}Gal 4}}%
      \csname LTb\endcsname
      \put(5816,940){\makebox(0,0)[r]{\strut{}Gal 5}}%
    }%
    \gplbacktext
    \put(0,0){\includegraphics{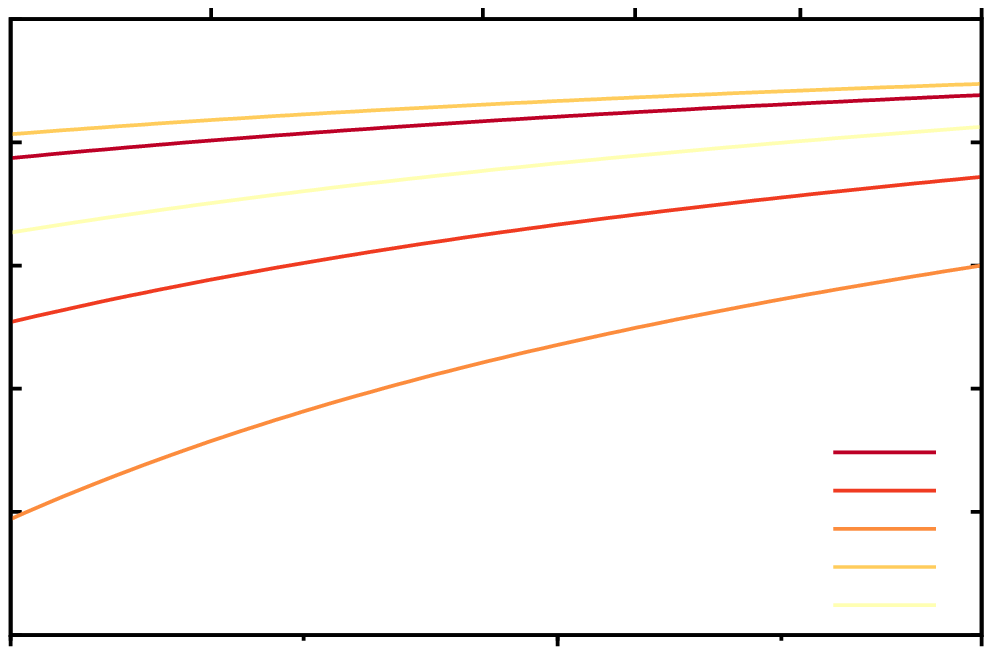}}%
    \gplfronttext
  \end{picture}%
\endgroup

%% file: perturbation.tex
\begingroup
  \makeatletter
  \providecommand\color[2][]{%
    \GenericError{(gnuplot) \space\space\space\@spaces}{%
      Package color not loaded in conjunction with
      terminal option `colourtext'%
    }{See the gnuplot documentation for explanation.%
    }{Either use 'blacktext' in gnuplot or load the package
      color.sty in LaTeX.}%
    \renewcommand\color[2][]{}%
  }%
  \providecommand\includegraphics[2][]{%
    \GenericError{(gnuplot) \space\space\space\@spaces}{%
      Package graphicx or graphics not loaded%
    }{See the gnuplot documentation for explanation.%
    }{The gnuplot epslatex terminal needs graphicx.sty or graphics.sty.}%
    \renewcommand\includegraphics[2][]{}%
  }%
  \providecommand\rotatebox[2]{#2}%
  \@ifundefined{ifGPcolor}{%
    \newif\ifGPcolor
    \GPcolortrue
  }{}%
  \@ifundefined{ifGPblacktext}{%
    \newif\ifGPblacktext
    \GPblacktexttrue
  }{}%
  \let\gplgaddtomacro\g@addto@macro
  \gdef\gplbacktext{}%
  \gdef\gplfronttext{}%
  \makeatother
  \ifGPblacktext
    \def\colorrgb#1{}%
    \def\colorgray#1{}%
  \else
    \ifGPcolor
      \def\colorrgb#1{\color[rgb]{#1}}%
      \def\colorgray#1{\color[gray]{#1}}%
      \expandafter\def\csname LTw\endcsname{\color{white}}%
      \expandafter\def\csname LTb\endcsname{\color{black}}%
      \expandafter\def\csname LTa\endcsname{\color{black}}%
      \expandafter\def\csname LT0\endcsname{\color[rgb]{1,0,0}}%
      \expandafter\def\csname LT1\endcsname{\color[rgb]{0,1,0}}%
      \expandafter\def\csname LT2\endcsname{\color[rgb]{0,0,1}}%
      \expandafter\def\csname LT3\endcsname{\color[rgb]{1,0,1}}%
      \expandafter\def\csname LT4\endcsname{\color[rgb]{0,1,1}}%
      \expandafter\def\csname LT5\endcsname{\color[rgb]{1,1,0}}%
      \expandafter\def\csname LT6\endcsname{\color[rgb]{0,0,0}}%
      \expandafter\def\csname LT7\endcsname{\color[rgb]{1,0.3,0}}%
      \expandafter\def\csname LT8\endcsname{\color[rgb]{0.5,0.5,0.5}}%
    \else
      \def\colorrgb#1{\color{black}}%
      \def\colorgray#1{\color[gray]{#1}}%
      \expandafter\def\csname LTw\endcsname{\color{white}}%
      \expandafter\def\csname LTb\endcsname{\color{black}}%
      \expandafter\def\csname LTa\endcsname{\color{black}}%
      \expandafter\def\csname LT0\endcsname{\color{black}}%
      \expandafter\def\csname LT1\endcsname{\color{black}}%
      \expandafter\def\csname LT2\endcsname{\color{black}}%
      \expandafter\def\csname LT3\endcsname{\color{black}}%
      \expandafter\def\csname LT4\endcsname{\color{black}}%
      \expandafter\def\csname LT5\endcsname{\color{black}}%
      \expandafter\def\csname LT6\endcsname{\color{black}}%
      \expandafter\def\csname LT7\endcsname{\color{black}}%
      \expandafter\def\csname LT8\endcsname{\color{black}}%
    \fi
  \fi
    \setlength{\unitlength}{0.0500bp}%
    \ifx\gptboxheight\undefined%
      \newlength{\gptboxheight}%
      \newlength{\gptboxwidth}%
      \newsavebox{\gptboxtext}%
    \fi%
    \setlength{\fboxrule}{0.5pt}%
    \setlength{\fboxsep}{1pt}%
\begin{picture}(7200.00,5040.00)%
    \gplgaddtomacro\gplbacktext{%
      \csname LTb\endcsname
      \put(814,704){\makebox(0,0)[r]{\strut{}$1.2$}}%
      \put(814,1161){\makebox(0,0)[r]{\strut{}$1.3$}}%
      \put(814,1618){\makebox(0,0)[r]{\strut{}$1.4$}}%
      \put(814,2076){\makebox(0,0)[r]{\strut{}$1.5$}}%
      \put(814,2533){\makebox(0,0)[r]{\strut{}$1.6$}}%
      \put(814,2990){\makebox(0,0)[r]{\strut{}$1.7$}}%
      \put(814,3447){\makebox(0,0)[r]{\strut{}$1.8$}}%
      \put(814,3905){\makebox(0,0)[r]{\strut{}$1.9$}}%
      \put(814,4362){\makebox(0,0)[r]{\strut{}$2$}}%
      \put(814,4819){\makebox(0,0)[r]{\strut{}$2.1$}}%
      \put(946,484){\makebox(0,0){\strut{}$0.01$}}%
      \put(3875,484){\makebox(0,0){\strut{}$0.1$}}%
      \put(6803,484){\makebox(0,0){\strut{}$1$}}%
    }%
    \gplgaddtomacro\gplfronttext{%
      \csname LTb\endcsname
      \put(198,2761){\rotatebox{-270}{\makebox(0,0){\strut{}$\lambda$}}}%
      \put(3874,154){\makebox(0,0){\strut{}$a$}}%
      \csname LTb\endcsname
      \put(1738,1757){\makebox(0,0)[r]{\strut{}Gal 1}}%
      \csname LTb\endcsname
      \put(1738,1537){\makebox(0,0)[r]{\strut{}Gal 2}}%
      \csname LTb\endcsname
      \put(1738,1317){\makebox(0,0)[r]{\strut{}Gal 3}}%
      \csname LTb\endcsname
      \put(1738,1097){\makebox(0,0)[r]{\strut{}Gal 4}}%
      \csname LTb\endcsname
      \put(1738,877){\makebox(0,0)[r]{\strut{}Gal 5}}%
    }%
    \gplbacktext
    \put(0,0){\includegraphics{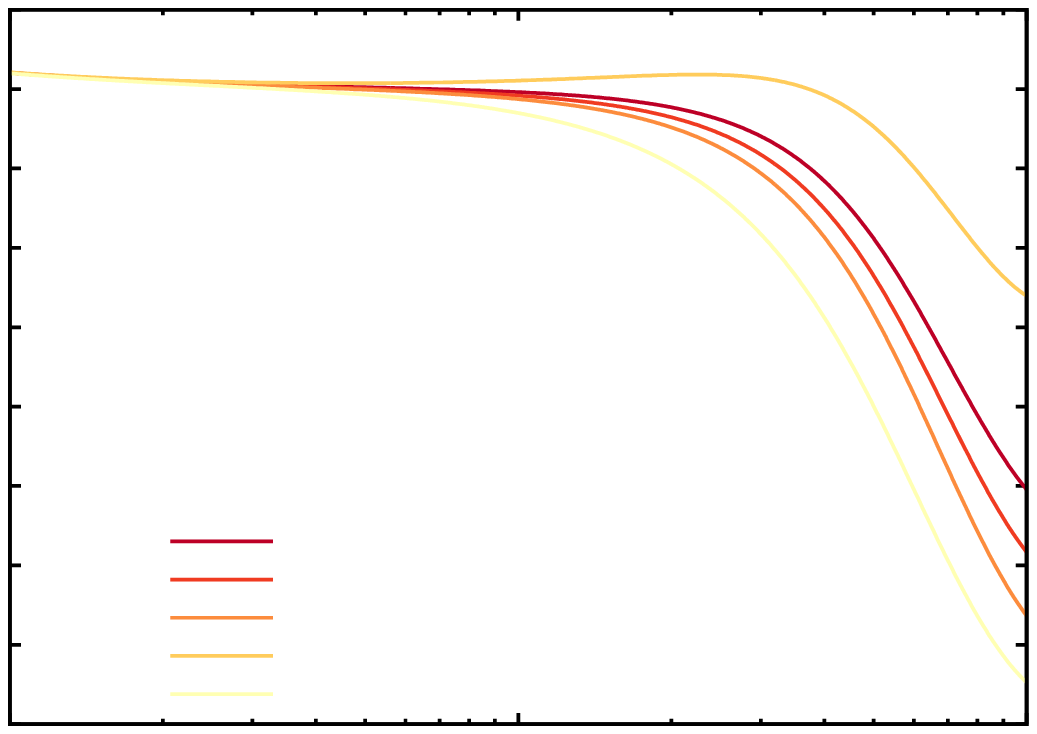}}%
    \gplfronttext
  \end{picture}%
\endgroup

%% file: growth_decay.tex
\begingroup
  \makeatletter
  \providecommand\color[2][]{%
    \GenericError{(gnuplot) \space\space\space\@spaces}{%
      Package color not loaded in conjunction with
      terminal option `colourtext'%
    }{See the gnuplot documentation for explanation.%
    }{Either use 'blacktext' in gnuplot or load the package
      color.sty in LaTeX.}%
    \renewcommand\color[2][]{}%
  }%
  \providecommand\includegraphics[2][]{%
    \GenericError{(gnuplot) \space\space\space\@spaces}{%
      Package graphicx or graphics not loaded%
    }{See the gnuplot documentation for explanation.%
    }{The gnuplot epslatex terminal needs graphicx.sty or graphics.sty.}%
    \renewcommand\includegraphics[2][]{}%
  }%
  \providecommand\rotatebox[2]{#2}%
  \@ifundefined{ifGPcolor}{%
    \newif\ifGPcolor
    \GPcolortrue
  }{}%
  \@ifundefined{ifGPblacktext}{%
    \newif\ifGPblacktext
    \GPblacktexttrue
  }{}%
  \let\gplgaddtomacro\g@addto@macro
  \gdef\gplbacktext{}%
  \gdef\gplfronttext{}%
  \makeatother
  \ifGPblacktext
    \def\colorrgb#1{}%
    \def\colorgray#1{}%
  \else
    \ifGPcolor
      \def\colorrgb#1{\color[rgb]{#1}}%
      \def\colorgray#1{\color[gray]{#1}}%
      \expandafter\def\csname LTw\endcsname{\color{white}}%
      \expandafter\def\csname LTb\endcsname{\color{black}}%
      \expandafter\def\csname LTa\endcsname{\color{black}}%
      \expandafter\def\csname LT0\endcsname{\color[rgb]{1,0,0}}%
      \expandafter\def\csname LT1\endcsname{\color[rgb]{0,1,0}}%
      \expandafter\def\csname LT2\endcsname{\color[rgb]{0,0,1}}%
      \expandafter\def\csname LT3\endcsname{\color[rgb]{1,0,1}}%
      \expandafter\def\csname LT4\endcsname{\color[rgb]{0,1,1}}%
      \expandafter\def\csname LT5\endcsname{\color[rgb]{1,1,0}}%
      \expandafter\def\csname LT6\endcsname{\color[rgb]{0,0,0}}%
      \expandafter\def\csname LT7\endcsname{\color[rgb]{1,0.3,0}}%
      \expandafter\def\csname LT8\endcsname{\color[rgb]{0.5,0.5,0.5}}%
    \else
      \def\colorrgb#1{\color{black}}%
      \def\colorgray#1{\color[gray]{#1}}%
      \expandafter\def\csname LTw\endcsname{\color{white}}%
      \expandafter\def\csname LTb\endcsname{\color{black}}%
      \expandafter\def\csname LTa\endcsname{\color{black}}%
      \expandafter\def\csname LT0\endcsname{\color{black}}%
      \expandafter\def\csname LT1\endcsname{\color{black}}%
      \expandafter\def\csname LT2\endcsname{\color{black}}%
      \expandafter\def\csname LT3\endcsname{\color{black}}%
      \expandafter\def\csname LT4\endcsname{\color{black}}%
      \expandafter\def\csname LT5\endcsname{\color{black}}%
      \expandafter\def\csname LT6\endcsname{\color{black}}%
      \expandafter\def\csname LT7\endcsname{\color{black}}%
      \expandafter\def\csname LT8\endcsname{\color{black}}%
    \fi
  \fi
    \setlength{\unitlength}{0.0500bp}%
    \ifx\gptboxheight\undefined%
      \newlength{\gptboxheight}%
      \newlength{\gptboxwidth}%
      \newsavebox{\gptboxtext}%
    \fi%
    \setlength{\fboxrule}{0.5pt}%
    \setlength{\fboxsep}{1pt}%
\begin{picture}(7200.00,5040.00)%
    \gplgaddtomacro\gplbacktext{%
      \csname LTb\endcsname
      \put(1078,704){\makebox(0,0)[r]{\strut{}$1\mathrm{e}{-9}$}}%
      \put(1078,1161){\makebox(0,0)[r]{\strut{}$1\mathrm{e}{-8}$}}%
      \put(1078,1618){\makebox(0,0)[r]{\strut{}$1\mathrm{e}{-7}$}}%
      \put(1078,2076){\makebox(0,0)[r]{\strut{}$1\mathrm{e}{-6}$}}%
      \put(1078,2533){\makebox(0,0)[r]{\strut{}$1\mathrm{e}{-5}$}}%
      \put(1078,2990){\makebox(0,0)[r]{\strut{}$1\mathrm{e}{-4}$}}%
      \put(1078,3447){\makebox(0,0)[r]{\strut{}$1\mathrm{e}{-3}$}}%
      \put(1078,3905){\makebox(0,0)[r]{\strut{}$1\mathrm{e}{-2}$}}%
      \put(1078,4362){\makebox(0,0)[r]{\strut{}$1\mathrm{e}{-1}$}}%
      \put(1078,4819){\makebox(0,0)[r]{\strut{}$1\mathrm{e}{0}$}}%
      \put(1595,484){\makebox(0,0){\strut{}$1\mathrm{e}{-3}$}}%
      \put(3331,484){\makebox(0,0){\strut{}$1\mathrm{e}{-2}$}}%
      \put(5067,484){\makebox(0,0){\strut{}$1\mathrm{e}{-1}$}}%
      \put(6803,484){\makebox(0,0){\strut{}$1\mathrm{e}{0}$}}%
    }%
    \gplgaddtomacro\gplfronttext{%
      \csname LTb\endcsname
      \put(198,2761){\rotatebox{-270}{\makebox(0,0){\strut{}$D_1(a)$, $D_2(a)$}}}%
      \put(4006,154){\makebox(0,0){\strut{}$a$}}%
      \csname LTb\endcsname
      \put(5816,2871){\makebox(0,0)[r]{\strut{}$D_1$}}%
      \csname LTb\endcsname
      \put(5816,2651){\makebox(0,0)[r]{\strut{}$D_2$}}%
    }%
    \gplbacktext
    \put(0,0){\includegraphics{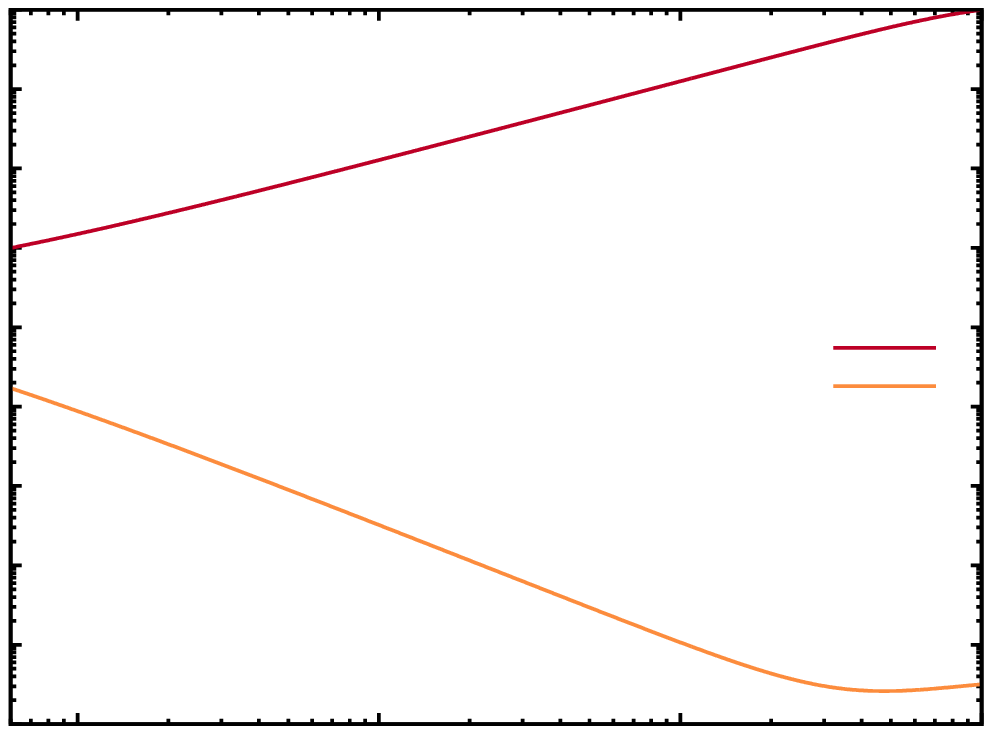}}%
    \gplfronttext
  \end{picture}%
\endgroup

%% file: growth_function.tex
\begingroup
  \makeatletter
  \providecommand\color[2][]{%
    \GenericError{(gnuplot) \space\space\space\@spaces}{%
      Package color not loaded in conjunction with
      terminal option `colourtext'%
    }{See the gnuplot documentation for explanation.%
    }{Either use 'blacktext' in gnuplot or load the package
      color.sty in LaTeX.}%
    \renewcommand\color[2][]{}%
  }%
  \providecommand\includegraphics[2][]{%
    \GenericError{(gnuplot) \space\space\space\@spaces}{%
      Package graphicx or graphics not loaded%
    }{See the gnuplot documentation for explanation.%
    }{The gnuplot epslatex terminal needs graphicx.sty or graphics.sty.}%
    \renewcommand\includegraphics[2][]{}%
  }%
  \providecommand\rotatebox[2]{#2}%
  \@ifundefined{ifGPcolor}{%
    \newif\ifGPcolor
    \GPcolortrue
  }{}%
  \@ifundefined{ifGPblacktext}{%
    \newif\ifGPblacktext
    \GPblacktexttrue
  }{}%
  \let\gplgaddtomacro\g@addto@macro
  \gdef\gplbacktext{}%
  \gdef\gplfronttext{}%
  \makeatother
  \ifGPblacktext
    \def\colorrgb#1{}%
    \def\colorgray#1{}%
  \else
    \ifGPcolor
      \def\colorrgb#1{\color[rgb]{#1}}%
      \def\colorgray#1{\color[gray]{#1}}%
      \expandafter\def\csname LTw\endcsname{\color{white}}%
      \expandafter\def\csname LTb\endcsname{\color{black}}%
      \expandafter\def\csname LTa\endcsname{\color{black}}%
      \expandafter\def\csname LT0\endcsname{\color[rgb]{1,0,0}}%
      \expandafter\def\csname LT1\endcsname{\color[rgb]{0,1,0}}%
      \expandafter\def\csname LT2\endcsname{\color[rgb]{0,0,1}}%
      \expandafter\def\csname LT3\endcsname{\color[rgb]{1,0,1}}%
      \expandafter\def\csname LT4\endcsname{\color[rgb]{0,1,1}}%
      \expandafter\def\csname LT5\endcsname{\color[rgb]{1,1,0}}%
      \expandafter\def\csname LT6\endcsname{\color[rgb]{0,0,0}}%
      \expandafter\def\csname LT7\endcsname{\color[rgb]{1,0.3,0}}%
      \expandafter\def\csname LT8\endcsname{\color[rgb]{0.5,0.5,0.5}}%
    \else
      \def\colorrgb#1{\color{black}}%
      \def\colorgray#1{\color[gray]{#1}}%
      \expandafter\def\csname LTw\endcsname{\color{white}}%
      \expandafter\def\csname LTb\endcsname{\color{black}}%
      \expandafter\def\csname LTa\endcsname{\color{black}}%
      \expandafter\def\csname LT0\endcsname{\color{black}}%
      \expandafter\def\csname LT1\endcsname{\color{black}}%
      \expandafter\def\csname LT2\endcsname{\color{black}}%
      \expandafter\def\csname LT3\endcsname{\color{black}}%
      \expandafter\def\csname LT4\endcsname{\color{black}}%
      \expandafter\def\csname LT5\endcsname{\color{black}}%
      \expandafter\def\csname LT6\endcsname{\color{black}}%
      \expandafter\def\csname LT7\endcsname{\color{black}}%
      \expandafter\def\csname LT8\endcsname{\color{black}}%
    \fi
  \fi
    \setlength{\unitlength}{0.0500bp}%
    \ifx\gptboxheight\undefined%
      \newlength{\gptboxheight}%
      \newlength{\gptboxwidth}%
      \newsavebox{\gptboxtext}%
    \fi%
    \setlength{\fboxrule}{0.5pt}%
    \setlength{\fboxsep}{1pt}%
\begin{picture}(7200.00,5040.00)%
    \gplgaddtomacro\gplbacktext{%
      \csname LTb\endcsname
      \put(1078,767){\makebox(0,0)[r]{\strut{}$0.001$}}%
      \put(1078,2542){\makebox(0,0)[r]{\strut{}$0.01$}}%
      \put(1078,4316){\makebox(0,0)[r]{\strut{}$0.1$}}%
      \put(1210,484){\makebox(0,0){\strut{}$0.1$}}%
      \put(6803,484){\makebox(0,0){\strut{}$1$}}%
      \put(1210,4599){\makebox(0,0){\strut{}$9$}}%
      \put(1466,4599){\makebox(0,0){\strut{}$8$}}%
      \put(1752,4599){\makebox(0,0){\strut{}$7$}}%
      \put(2076,4599){\makebox(0,0){\strut{}$6$}}%
      \put(2451,4599){\makebox(0,0){\strut{}$5$}}%
      \put(2894,4599){\makebox(0,0){\strut{}$4$}}%
      \put(3436,4599){\makebox(0,0){\strut{}$3$}}%
      \put(4134,4599){\makebox(0,0){\strut{}$2$}}%
      \put(5119,4599){\makebox(0,0){\strut{}$1$}}%
      \put(6803,4599){\makebox(0,0){\strut{}$0$}}%
    }%
    \gplgaddtomacro\gplfronttext{%
      \csname LTb\endcsname
      \put(198,2541){\rotatebox{-270}{\makebox(0,0){\strut{}$f^{\rm (Gal)}/f^{(\Lambda)}-1$}}}%
      \put(4006,154){\makebox(0,0){\strut{}$a$}}%
      \put(4006,4929){\makebox(0,0){\strut{}$z$}}%
      \csname LTb\endcsname
      \put(2002,4143){\makebox(0,0)[r]{\strut{}Gal 1}}%
      \csname LTb\endcsname
      \put(2002,3923){\makebox(0,0)[r]{\strut{}Gal 2}}%
      \csname LTb\endcsname
      \put(2002,3703){\makebox(0,0)[r]{\strut{}Gal 3}}%
      \csname LTb\endcsname
      \put(2002,3483){\makebox(0,0)[r]{\strut{}Gal 4}}%
      \csname LTb\endcsname
      \put(2002,3263){\makebox(0,0)[r]{\strut{}Gal 5}}%
    }%
    \gplbacktext
    \put(0,0){\includegraphics{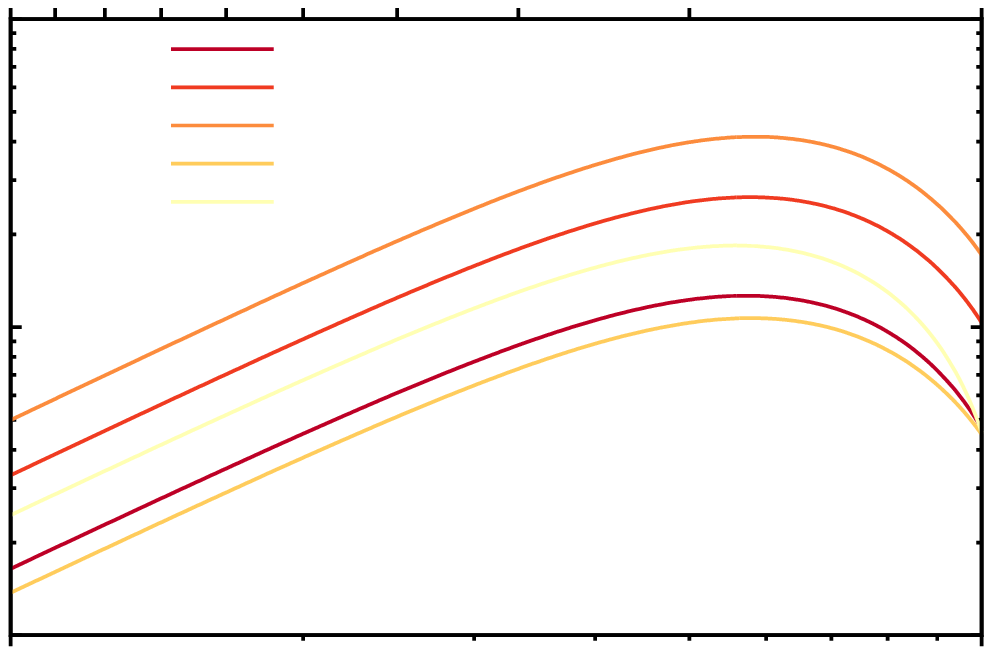}}%
    \gplfronttext
  \end{picture}%
\endgroup

%% file: bispectrum.tex
\begingroup
  \makeatletter
  \providecommand\color[2][]{%
    \GenericError{(gnuplot) \space\space\space\@spaces}{%
      Package color not loaded in conjunction with
      terminal option `colourtext'%
    }{See the gnuplot documentation for explanation.%
    }{Either use 'blacktext' in gnuplot or load the package
      color.sty in LaTeX.}%
    \renewcommand\color[2][]{}%
  }%
  \providecommand\includegraphics[2][]{%
    \GenericError{(gnuplot) \space\space\space\@spaces}{%
      Package graphicx or graphics not loaded%
    }{See the gnuplot documentation for explanation.%
    }{The gnuplot epslatex terminal needs graphicx.sty or graphics.sty.}%
    \renewcommand\includegraphics[2][]{}%
  }%
  \providecommand\rotatebox[2]{#2}%
  \@ifundefined{ifGPcolor}{%
    \newif\ifGPcolor
    \GPcolortrue
  }{}%
  \@ifundefined{ifGPblacktext}{%
    \newif\ifGPblacktext
    \GPblacktexttrue
  }{}%
  \let\gplgaddtomacro\g@addto@macro
  \gdef\gplbacktext{}%
  \gdef\gplfronttext{}%
  \makeatother
  \ifGPblacktext
    \def\colorrgb#1{}%
    \def\colorgray#1{}%
  \else
    \ifGPcolor
      \def\colorrgb#1{\color[rgb]{#1}}%
      \def\colorgray#1{\color[gray]{#1}}%
      \expandafter\def\csname LTw\endcsname{\color{white}}%
      \expandafter\def\csname LTb\endcsname{\color{black}}%
      \expandafter\def\csname LTa\endcsname{\color{black}}%
      \expandafter\def\csname LT0\endcsname{\color[rgb]{1,0,0}}%
      \expandafter\def\csname LT1\endcsname{\color[rgb]{0,1,0}}%
      \expandafter\def\csname LT2\endcsname{\color[rgb]{0,0,1}}%
      \expandafter\def\csname LT3\endcsname{\color[rgb]{1,0,1}}%
      \expandafter\def\csname LT4\endcsname{\color[rgb]{0,1,1}}%
      \expandafter\def\csname LT5\endcsname{\color[rgb]{1,1,0}}%
      \expandafter\def\csname LT6\endcsname{\color[rgb]{0,0,0}}%
      \expandafter\def\csname LT7\endcsname{\color[rgb]{1,0.3,0}}%
      \expandafter\def\csname LT8\endcsname{\color[rgb]{0.5,0.5,0.5}}%
    \else
      \def\colorrgb#1{\color{black}}%
      \def\colorgray#1{\color[gray]{#1}}%
      \expandafter\def\csname LTw\endcsname{\color{white}}%
      \expandafter\def\csname LTb\endcsname{\color{black}}%
      \expandafter\def\csname LTa\endcsname{\color{black}}%
      \expandafter\def\csname LT0\endcsname{\color{black}}%
      \expandafter\def\csname LT1\endcsname{\color{black}}%
      \expandafter\def\csname LT2\endcsname{\color{black}}%
      \expandafter\def\csname LT3\endcsname{\color{black}}%
      \expandafter\def\csname LT4\endcsname{\color{black}}%
      \expandafter\def\csname LT5\endcsname{\color{black}}%
      \expandafter\def\csname LT6\endcsname{\color{black}}%
      \expandafter\def\csname LT7\endcsname{\color{black}}%
      \expandafter\def\csname LT8\endcsname{\color{black}}%
    \fi
  \fi
    \setlength{\unitlength}{0.0500bp}%
    \ifx\gptboxheight\undefined%
      \newlength{\gptboxheight}%
      \newlength{\gptboxwidth}%
      \newsavebox{\gptboxtext}%
    \fi%
    \setlength{\fboxrule}{0.5pt}%
    \setlength{\fboxsep}{1pt}%
\begin{picture}(7200.00,5040.00)%
    \gplgaddtomacro\gplbacktext{%
      \csname LTb\endcsname
      \put(1078,1907){\makebox(0,0)[r]{\strut{}$0.001$}}%
      \put(1078,3538){\makebox(0,0)[r]{\strut{}$0.01$}}%
      \put(1210,484){\makebox(0,0){\strut{}$0.1$}}%
      \put(6803,484){\makebox(0,0){\strut{}$1$}}%
      \put(1210,4599){\makebox(0,0){\strut{}$9$}}%
      \put(1466,4599){\makebox(0,0){\strut{}$8$}}%
      \put(1752,4599){\makebox(0,0){\strut{}$7$}}%
      \put(2076,4599){\makebox(0,0){\strut{}$6$}}%
      \put(2451,4599){\makebox(0,0){\strut{}$5$}}%
      \put(2894,4599){\makebox(0,0){\strut{}$4$}}%
      \put(3436,4599){\makebox(0,0){\strut{}$3$}}%
      \put(4134,4599){\makebox(0,0){\strut{}$2$}}%
      \put(5119,4599){\makebox(0,0){\strut{}$1$}}%
      \put(6803,4599){\makebox(0,0){\strut{}$0$}}%
    }%
    \gplgaddtomacro\gplfronttext{%
      \csname LTb\endcsname
      \put(198,2541){\rotatebox{-270}{\makebox(0,0){\strut{}$\mathcal{B}^{\rm (Gal)}/\mathcal{B}^{(\Lambda)}-1$}}}%
      \put(4006,154){\makebox(0,0){\strut{}$a$}}%
      \put(4006,4929){\makebox(0,0){\strut{}$z$}}%
      \csname LTb\endcsname
      \put(1804,4143){\makebox(0,0)[r]{\strut{}Gal 1}}%
      \csname LTb\endcsname
      \put(1804,3923){\makebox(0,0)[r]{\strut{}Gal 2}}%
      \csname LTb\endcsname
      \put(1804,3703){\makebox(0,0)[r]{\strut{}Gal 3}}%
      \csname LTb\endcsname
      \put(1804,3483){\makebox(0,0)[r]{\strut{}Gal 4}}%
      \csname LTb\endcsname
      \put(1804,3263){\makebox(0,0)[r]{\strut{}Gal 5}}%
    }%
    \gplbacktext
    \put(0,0){\includegraphics{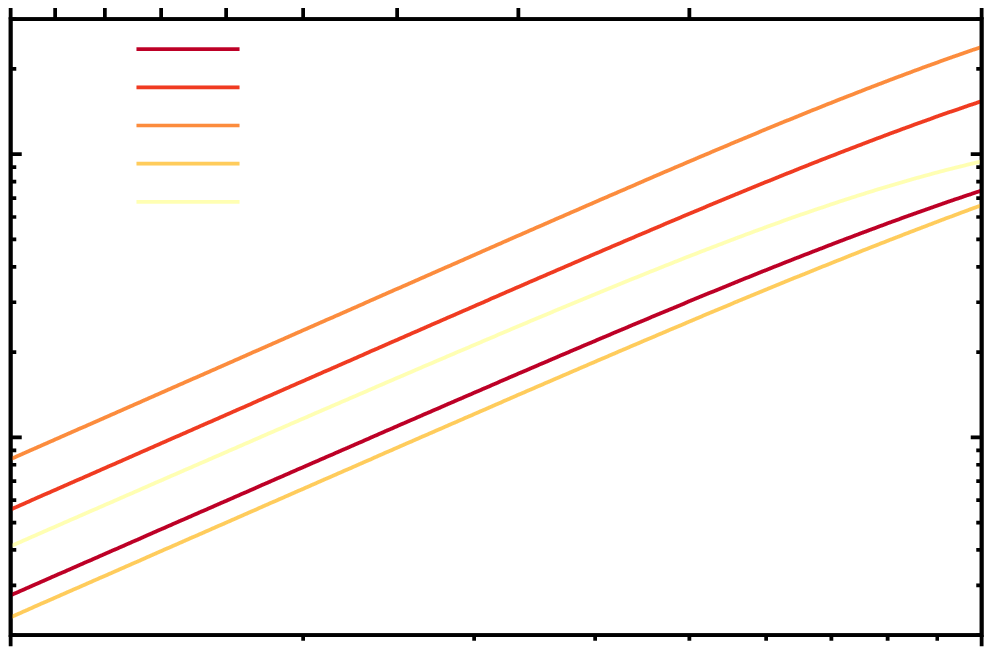}}%
    \gplfronttext
  \end{picture}%
\endgroup

%% file: curly_B.tex
\begingroup
  \makeatletter
  \providecommand\color[2][]{%
    \GenericError{(gnuplot) \space\space\space\@spaces}{%
      Package color not loaded in conjunction with
      terminal option `colourtext'%
    }{See the gnuplot documentation for explanation.%
    }{Either use 'blacktext' in gnuplot or load the package
      color.sty in LaTeX.}%
    \renewcommand\color[2][]{}%
  }%
  \providecommand\includegraphics[2][]{%
    \GenericError{(gnuplot) \space\space\space\@spaces}{%
      Package graphicx or graphics not loaded%
    }{See the gnuplot documentation for explanation.%
    }{The gnuplot epslatex terminal needs graphicx.sty or graphics.sty.}%
    \renewcommand\includegraphics[2][]{}%
  }%
  \providecommand\rotatebox[2]{#2}%
  \@ifundefined{ifGPcolor}{%
    \newif\ifGPcolor
    \GPcolortrue
  }{}%
  \@ifundefined{ifGPblacktext}{%
    \newif\ifGPblacktext
    \GPblacktexttrue
  }{}%
  \let\gplgaddtomacro\g@addto@macro
  \gdef\gplbacktext{}%
  \gdef\gplfronttext{}%
  \makeatother
  \ifGPblacktext
    \def\colorrgb#1{}%
    \def\colorgray#1{}%
  \else
    \ifGPcolor
      \def\colorrgb#1{\color[rgb]{#1}}%
      \def\colorgray#1{\color[gray]{#1}}%
      \expandafter\def\csname LTw\endcsname{\color{white}}%
      \expandafter\def\csname LTb\endcsname{\color{black}}%
      \expandafter\def\csname LTa\endcsname{\color{black}}%
      \expandafter\def\csname LT0\endcsname{\color[rgb]{1,0,0}}%
      \expandafter\def\csname LT1\endcsname{\color[rgb]{0,1,0}}%
      \expandafter\def\csname LT2\endcsname{\color[rgb]{0,0,1}}%
      \expandafter\def\csname LT3\endcsname{\color[rgb]{1,0,1}}%
      \expandafter\def\csname LT4\endcsname{\color[rgb]{0,1,1}}%
      \expandafter\def\csname LT5\endcsname{\color[rgb]{1,1,0}}%
      \expandafter\def\csname LT6\endcsname{\color[rgb]{0,0,0}}%
      \expandafter\def\csname LT7\endcsname{\color[rgb]{1,0.3,0}}%
      \expandafter\def\csname LT8\endcsname{\color[rgb]{0.5,0.5,0.5}}%
    \else
      \def\colorrgb#1{\color{black}}%
      \def\colorgray#1{\color[gray]{#1}}%
      \expandafter\def\csname LTw\endcsname{\color{white}}%
      \expandafter\def\csname LTb\endcsname{\color{black}}%
      \expandafter\def\csname LTa\endcsname{\color{black}}%
      \expandafter\def\csname LT0\endcsname{\color{black}}%
      \expandafter\def\csname LT1\endcsname{\color{black}}%
      \expandafter\def\csname LT2\endcsname{\color{black}}%
      \expandafter\def\csname LT3\endcsname{\color{black}}%
      \expandafter\def\csname LT4\endcsname{\color{black}}%
      \expandafter\def\csname LT5\endcsname{\color{black}}%
      \expandafter\def\csname LT6\endcsname{\color{black}}%
      \expandafter\def\csname LT7\endcsname{\color{black}}%
      \expandafter\def\csname LT8\endcsname{\color{black}}%
    \fi
  \fi
    \setlength{\unitlength}{0.0500bp}%
    \ifx\gptboxheight\undefined%
      \newlength{\gptboxheight}%
      \newlength{\gptboxwidth}%
      \newsavebox{\gptboxtext}%
    \fi%
    \setlength{\fboxrule}{0.5pt}%
    \setlength{\fboxsep}{1pt}%
\begin{picture}(7200.00,5040.00)%
    \gplgaddtomacro\gplbacktext{%
      \csname LTb\endcsname
      \put(1078,1079){\makebox(0,0)[r]{\strut{}$1\mathrm{e}{-4}$}}%
      \put(1078,2486){\makebox(0,0)[r]{\strut{}$1\mathrm{e}{-3}$}}%
      \put(1078,3893){\makebox(0,0)[r]{\strut{}$1\mathrm{e}{-2}$}}%
      \put(1210,484){\makebox(0,0){\strut{}$0.1$}}%
      \put(6803,484){\makebox(0,0){\strut{}$1$}}%
      \put(1210,4599){\makebox(0,0){\strut{}$9$}}%
      \put(1466,4599){\makebox(0,0){\strut{}$8$}}%
      \put(1752,4599){\makebox(0,0){\strut{}$7$}}%
      \put(2076,4599){\makebox(0,0){\strut{}$6$}}%
      \put(2451,4599){\makebox(0,0){\strut{}$5$}}%
      \put(2894,4599){\makebox(0,0){\strut{}$4$}}%
      \put(3436,4599){\makebox(0,0){\strut{}$3$}}%
      \put(4134,4599){\makebox(0,0){\strut{}$2$}}%
      \put(5119,4599){\makebox(0,0){\strut{}$1$}}%
      \put(6803,4599){\makebox(0,0){\strut{}$0$}}%
    }%
    \gplgaddtomacro\gplfronttext{%
      \csname LTb\endcsname
      \put(198,2541){\rotatebox{-270}{\makebox(0,0){\strut{}$\mathcal{B}$}}}%
      \put(4006,154){\makebox(0,0){\strut{}$a$}}%
      \put(4006,4929){\makebox(0,0){\strut{}$z$}}%
      \csname LTb\endcsname
      \put(1870,4143){\makebox(0,0)[r]{\strut{}Gal 1}}%
      \csname LTb\endcsname
      \put(1870,3923){\makebox(0,0)[r]{\strut{}Gal 2}}%
      \csname LTb\endcsname
      \put(1870,3703){\makebox(0,0)[r]{\strut{}Gal 3}}%
      \csname LTb\endcsname
      \put(1870,3483){\makebox(0,0)[r]{\strut{}Gal 4}}%
      \csname LTb\endcsname
      \put(1870,3263){\makebox(0,0)[r]{\strut{}Gal 5}}%
    }%
    \gplgaddtomacro\gplbacktext{%
    }%
    \gplgaddtomacro\gplfronttext{%
      \csname LTb\endcsname
      \put(4525,4096){\makebox(0,0)[r]{\strut{}$-\mathcal{B}_\pi$}}%
      \csname LTb\endcsname
      \put(4525,3876){\makebox(0,0)[r]{\strut{}$\mathcal{B}_{GR}^{\rm (Gal)}-\mathcal{B}_{GR}^{(\Lambda)}$}}%
    }%
    \gplbacktext
    \put(0,0){\includegraphics{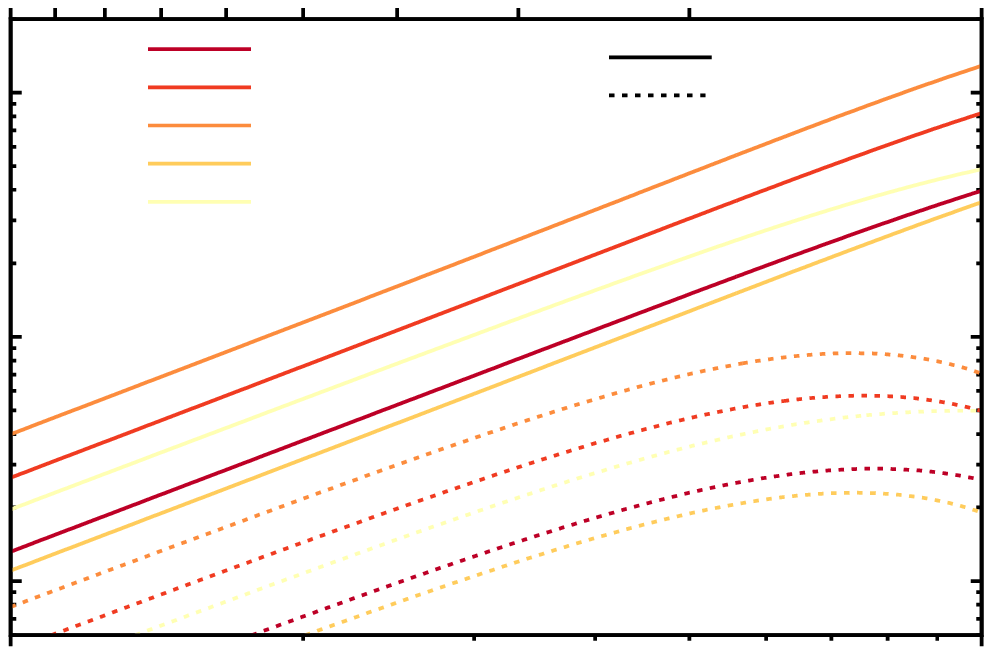}}%
    \gplfronttext
  \end{picture}%
\endgroup

%% file: red_bispec.tex
\begingroup
  \makeatletter
  \providecommand\color[2][]{%
    \GenericError{(gnuplot) \space\space\space\@spaces}{%
      Package color not loaded in conjunction with
      terminal option `colourtext'%
    }{See the gnuplot documentation for explanation.%
    }{Either use 'blacktext' in gnuplot or load the package
      color.sty in LaTeX.}%
    \renewcommand\color[2][]{}%
  }%
  \providecommand\includegraphics[2][]{%
    \GenericError{(gnuplot) \space\space\space\@spaces}{%
      Package graphicx or graphics not loaded%
    }{See the gnuplot documentation for explanation.%
    }{The gnuplot epslatex terminal needs graphicx.sty or graphics.sty.}%
    \renewcommand\includegraphics[2][]{}%
  }%
  \providecommand\rotatebox[2]{#2}%
  \@ifundefined{ifGPcolor}{%
    \newif\ifGPcolor
    \GPcolortrue
  }{}%
  \@ifundefined{ifGPblacktext}{%
    \newif\ifGPblacktext
    \GPblacktexttrue
  }{}%
  \let\gplgaddtomacro\g@addto@macro
  \gdef\gplbacktext{}%
  \gdef\gplfronttext{}%
  \makeatother
  \ifGPblacktext
    \def\colorrgb#1{}%
    \def\colorgray#1{}%
  \else
    \ifGPcolor
      \def\colorrgb#1{\color[rgb]{#1}}%
      \def\colorgray#1{\color[gray]{#1}}%
      \expandafter\def\csname LTw\endcsname{\color{white}}%
      \expandafter\def\csname LTb\endcsname{\color{black}}%
      \expandafter\def\csname LTa\endcsname{\color{black}}%
      \expandafter\def\csname LT0\endcsname{\color[rgb]{1,0,0}}%
      \expandafter\def\csname LT1\endcsname{\color[rgb]{0,1,0}}%
      \expandafter\def\csname LT2\endcsname{\color[rgb]{0,0,1}}%
      \expandafter\def\csname LT3\endcsname{\color[rgb]{1,0,1}}%
      \expandafter\def\csname LT4\endcsname{\color[rgb]{0,1,1}}%
      \expandafter\def\csname LT5\endcsname{\color[rgb]{1,1,0}}%
      \expandafter\def\csname LT6\endcsname{\color[rgb]{0,0,0}}%
      \expandafter\def\csname LT7\endcsname{\color[rgb]{1,0.3,0}}%
      \expandafter\def\csname LT8\endcsname{\color[rgb]{0.5,0.5,0.5}}%
    \else
      \def\colorrgb#1{\color{black}}%
      \def\colorgray#1{\color[gray]{#1}}%
      \expandafter\def\csname LTw\endcsname{\color{white}}%
      \expandafter\def\csname LTb\endcsname{\color{black}}%
      \expandafter\def\csname LTa\endcsname{\color{black}}%
      \expandafter\def\csname LT0\endcsname{\color{black}}%
      \expandafter\def\csname LT1\endcsname{\color{black}}%
      \expandafter\def\csname LT2\endcsname{\color{black}}%
      \expandafter\def\csname LT3\endcsname{\color{black}}%
      \expandafter\def\csname LT4\endcsname{\color{black}}%
      \expandafter\def\csname LT5\endcsname{\color{black}}%
      \expandafter\def\csname LT6\endcsname{\color{black}}%
      \expandafter\def\csname LT7\endcsname{\color{black}}%
      \expandafter\def\csname LT8\endcsname{\color{black}}%
    \fi
  \fi
    \setlength{\unitlength}{0.0500bp}%
    \ifx\gptboxheight\undefined%
      \newlength{\gptboxheight}%
      \newlength{\gptboxwidth}%
      \newsavebox{\gptboxtext}%
    \fi%
    \setlength{\fboxrule}{0.5pt}%
    \setlength{\fboxsep}{1pt}%
\begin{picture}(7200.00,5040.00)%
    \gplgaddtomacro\gplbacktext{%
      \csname LTb\endcsname
      \put(948,594){\makebox(0,0)[r]{\strut{}$-0.018$}}%
      \put(948,1027){\makebox(0,0)[r]{\strut{}$-0.016$}}%
      \put(948,1460){\makebox(0,0)[r]{\strut{}$-0.014$}}%
      \put(948,1892){\makebox(0,0)[r]{\strut{}$-0.012$}}%
      \put(948,2325){\makebox(0,0)[r]{\strut{}$-0.01$}}%
      \put(948,2758){\makebox(0,0)[r]{\strut{}$-0.008$}}%
      \put(948,3191){\makebox(0,0)[r]{\strut{}$-0.006$}}%
      \put(948,3623){\makebox(0,0)[r]{\strut{}$-0.004$}}%
      \put(948,4056){\makebox(0,0)[r]{\strut{}$-0.002$}}%
      \put(948,4489){\makebox(0,0)[r]{\strut{}$0$}}%
      \put(1211,374){\makebox(0,0){\strut{}$-1$}}%
      \put(1865,374){\makebox(0,0){\strut{}$-0.5$}}%
      \put(2520,374){\makebox(0,0){\strut{}$0$}}%
      \put(3174,374){\makebox(0,0){\strut{}$0.5$}}%
      \put(3828,374){\makebox(0,0){\strut{}$1$}}%
    }%
    \gplgaddtomacro\gplfronttext{%
      \csname LTb\endcsname
      \put(68,2541){\rotatebox{-270}{\makebox(0,0){\strut{}$Q^{\rm (Gal)}/Q^{(\Lambda)}-1$}}}%
      \put(2519,154){\makebox(0,0){\strut{}$\mu_{12}$}}%
      \put(2519,4709){\makebox(0,0){\strut{}$(a)$: $k_1=k_2=\SI{0.1}{Mpc/h}$}}%
      \csname LTb\endcsname
      \put(2972,1647){\makebox(0,0)[r]{\strut{}Gal 1}}%
      \csname LTb\endcsname
      \put(2972,1427){\makebox(0,0)[r]{\strut{}Gal 2}}%
      \csname LTb\endcsname
      \put(2972,1207){\makebox(0,0)[r]{\strut{}Gal 3}}%
      \csname LTb\endcsname
      \put(2972,987){\makebox(0,0)[r]{\strut{}Gal 4}}%
      \csname LTb\endcsname
      \put(2972,767){\makebox(0,0)[r]{\strut{}Gal 5}}%
    }%
    \gplgaddtomacro\gplbacktext{%
      \csname LTb\endcsname
      \put(3828,594){\makebox(0,0)[r]{\strut{}}}%
      \put(3828,1027){\makebox(0,0)[r]{\strut{}}}%
      \put(3828,1460){\makebox(0,0)[r]{\strut{}}}%
      \put(3828,1892){\makebox(0,0)[r]{\strut{}}}%
      \put(3828,2325){\makebox(0,0)[r]{\strut{}}}%
      \put(3828,2758){\makebox(0,0)[r]{\strut{}}}%
      \put(3828,3191){\makebox(0,0)[r]{\strut{}}}%
      \put(3828,3623){\makebox(0,0)[r]{\strut{}}}%
      \put(3828,4056){\makebox(0,0)[r]{\strut{}}}%
      \put(3828,4489){\makebox(0,0)[r]{\strut{}}}%
      \put(4091,374){\makebox(0,0){\strut{}$-1$}}%
      \put(4745,374){\makebox(0,0){\strut{}$-0.5$}}%
      \put(5400,374){\makebox(0,0){\strut{}$0$}}%
      \put(6054,374){\makebox(0,0){\strut{}$0.5$}}%
      \put(6708,374){\makebox(0,0){\strut{}$1$}}%
    }%
    \gplgaddtomacro\gplfronttext{%
      \csname LTb\endcsname
      \put(5399,154){\makebox(0,0){\strut{}$\mu_{12}$}}%
      \put(5399,4709){\makebox(0,0){\strut{}$(b)$: $k_1=2\times k_2=\SI{0.1}{Mpc/h}$}}%
      \csname LTb\endcsname
      \put(5852,1647){\makebox(0,0)[r]{\strut{}Gal 1}}%
      \csname LTb\endcsname
      \put(5852,1427){\makebox(0,0)[r]{\strut{}Gal 2}}%
      \csname LTb\endcsname
      \put(5852,1207){\makebox(0,0)[r]{\strut{}Gal 3}}%
      \csname LTb\endcsname
      \put(5852,987){\makebox(0,0)[r]{\strut{}Gal 4}}%
      \csname LTb\endcsname
      \put(5852,767){\makebox(0,0)[r]{\strut{}Gal 5}}%
    }%
    \gplbacktext
    \put(0,0){\includegraphics{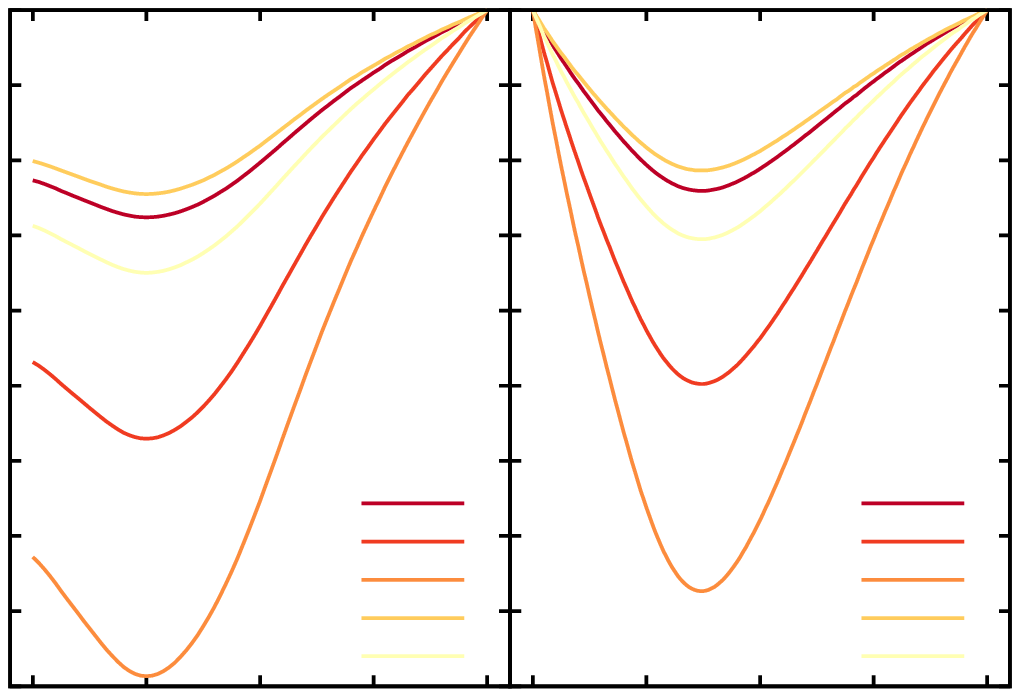}}%
    \gplfronttext
  \end{picture}%
\endgroup